# Are intermediate range periodicities in sunspot area associated with planetary motion?

Ian Edmonds
12 Lentara St, Kenmore, Brisbane, Australia 4069.
Ph/Fax 61 7 3378 6586, ian@solartran.com.au

**Abstract.**
Observations of periodicities in solar activity in the intermediate period range have been reported numerous times in the past 30 years. However, no accepted explanation for the occurrence of the quasi-periodicities has emerged. In this paper we examine the possibility that the periodicities can be linked to Mercury or to Mercury-planet conjunction periods. We show that peaks at the 45 day Mercury-Jupiter conjunction period, the 116 day first harmonic period of the Mercury-Earth conjunction and the 289 day period of the sub harmonic Mercury-Venus/Mercury-Earth conjunctions are prominent in spectra of sunspot area. We observe two prominent peaks close to the 88 day Mercury periodicity and two prominent peaks close to the 176 day first sub harmonic of Mercury periodicity. To confirm that the peaks arise as sidebands we band pass filter the sunspot area record to isolate the ~88 and ~176 day components of sunspot area. The components occur in episodes of duration from 1.5 to 4 years, with successive episodes usually overlapping in time but, for significant intervals in the record, the episodes were discrete, i.e. not overlapping. The time variation of the components was compared with the time variation of the orbital radius, $R_M$, of Mercury – or more specifically, with the time variation of the factor $(1/R_M)^3$, which is proportional to the tidal effect of Mercury. We were able to show that whenever episodes were discrete the time variation of the component of sunspot area during the episode was either in-phase or in anti-phase with the time variation of $(1/R_M)^3$. We interpret this as indicating a connection between planetary motion and sunspot emergence. When several discrete episodes of the components occurred during a solar cycle the spectrum of sunspot area during that solar cycle exhibited periodicities at sidebands to the 88 day or 176 day periods, the periodicity of the sidebands depending on the duration of the episodes. A simple model based on amplitude modulation of 88 day and 176 day period sinusoids was able to consistently reproduce the spectral peaks, or Rieger quasi-periodicities, observed in the spectra of sunspot area. Thus the seven most prominent peaks in the intermediate range of sunspot area periodicity could be attributed to Mercury-planet periodicity. It is proposed that the observed connection between planetary motion and sunspot emergence involves Rossby waves with mode periods at Mercury – planet periods and the triggering of sunspot emergence by those Rossby waves.

## 1. Introduction

A model accounting for the ~11 year cycle of sunspot emergence and decay was described by Babcock (1961). In this model solar magnetic flux in the interior becomes unstable and loops of flux float up to the surface to form bipolar sunspots. Opposite poles of the sunspots migrate to the solar poles, and eventually the sunspots are eliminated and the polar magnetic field reversed. This is the accepted explanation of the decadal variation in solar activity. However, it is not clear why the cycle is ~11 years long. Connection to the motions of Jupiter and Saturn has been suggested as an explanation, (Charbonneau 2002, Scafetta 2012, Charbonneau 2013). The other major periodicity observed in solar activity is the rotation period of the Sun which brings sunspots into view of Earth every 27 days.

Periodicities in solar activity are also observed in the intermediate period range between the ~11 year cycle and the ~ 27 day cycle. Intermediate periodicities are known as quasi-periodicities, Rieger (1984), due to their intermittency, varying periodicity and number. For example, Tan and Cheng (2013) report dozens of intermediate range periodicities. The quasi-periodicities occur in episodes of 1 – 3 years duration around the maximum of each solar cycle with periodicity around 150 - 160 days often observed, (Lean 1990, Richardson and Cane 2005, Ballester et al 2004). However the range of intermediate periodicity reported extends from 40 days to several years, (Chowdhury et al 2015, Tan and Chen 2013). Relevant to this paper are reports of intermediate periodicities in sunspot number and area, (Lean & Brueckner 1989, Lean 1990, Pap et al 1990, Carbonell & Ballester 1990, Carbonell & Ballester 1992, Verma and Joshi 1987, Verma et al 1992, Oliver and Ballester 1995, Oliver et al 1998, Ballester et al 1999, Krivova & Solanki 2002, Chowdhury et al 2009, Getko 2014 and Chowdhury et al 2015). Zaqarashvili et al (2010) reported time/period diagrams showing quasi-periodicities in sunspot area for cycles 19 to 23. The diagrams show a broad range of periods and indicate strong episodes of periodicity in the period range 150 to 190 days occurring around the maximum of solar cycles 19 and 21. Most recently Kolotkov et al (2015) studied the quasi-periodicities in sunspot area during the last two solar cycles identifying thirteen significant periodicities in the range between 25 and 4000 days. Currently, there is no accepted explanation for intermediate range quasi-periodicities.

Proposed explanations for quasi-periodicities include Bai (1987) suggesting that the cause must be a mechanism that induces active regions to emit flares; Ichimoto et al (1985) suggesting that the quasi-periodicities are associated with the storage and escape of magnetic flux from the Sun; Bai & Cliver (1990) suggesting the periodicities could be simulated with a non-linear damped oscillator; Bai & Sturrock (1991) and Sturrock & Bai (1993) suggesting that the Sun contains a "clock" with a period of 25.5 days and the periodicities are sub-harmonics of the "clock" period; Wolff (1983, 1992, 1998) suggesting generation via the interaction of the solar activity band with solar g modes; Sturrock (1996) suggesting the Sun contains two "rotators", one at ~22 days and the other at ~25 days, that combine to produce the observed periodicities; Seker (2012) suggesting resonance between Alfven waves and planetary tides; Lockwood (2001) suggesting variations of ~ 1 year duration may be related to oscillations close to the base of the solar convection zone; Wang and Sheeley (2003) demonstrating that periodicities in the intermediate range in solar magnetic flux can occur through the random development of sunspot groups on the solar surface; and Lou (2000), Lou et al (2003), Zaqarashvili et al (2010) and Gurgenashvili et al (2016) suggesting the periodicities in the intermediate range can be linked to equatorially trapped Rossby waves near the surface of the Sun.

Rossby waves can form in a fluid layer on the surface of a sphere. Lou (2000) and earlier Wolff (1998) derived dispersion relations for equatorially trapped Rossby waves in the photosphere of the Sun. The amplitude of equatorially trapped Rossby waves vary longitudinally around the solar equator and have a Gaussian envelope spanning 60° across the equator. Lou (2000) derived the following expression for the allowed periods:

$$T(p,q) = 25.1[q/2 + 0.17(2p + 1)/q] \text{ days} \tag{1}$$

where 25.1 days is the sidereal period of surface rotation of the Sun at the equator, $p = 1$, 2 and $q = p, p+1, p+2 \ldots$ The $p = 1$ mode has one nodal line through the pole, the $p = 2$ mode has two nodal lines through the poles. Figure 1 shows the allowed mode periods for the $p = 1$ and $p = 2$ modes as a function of q.

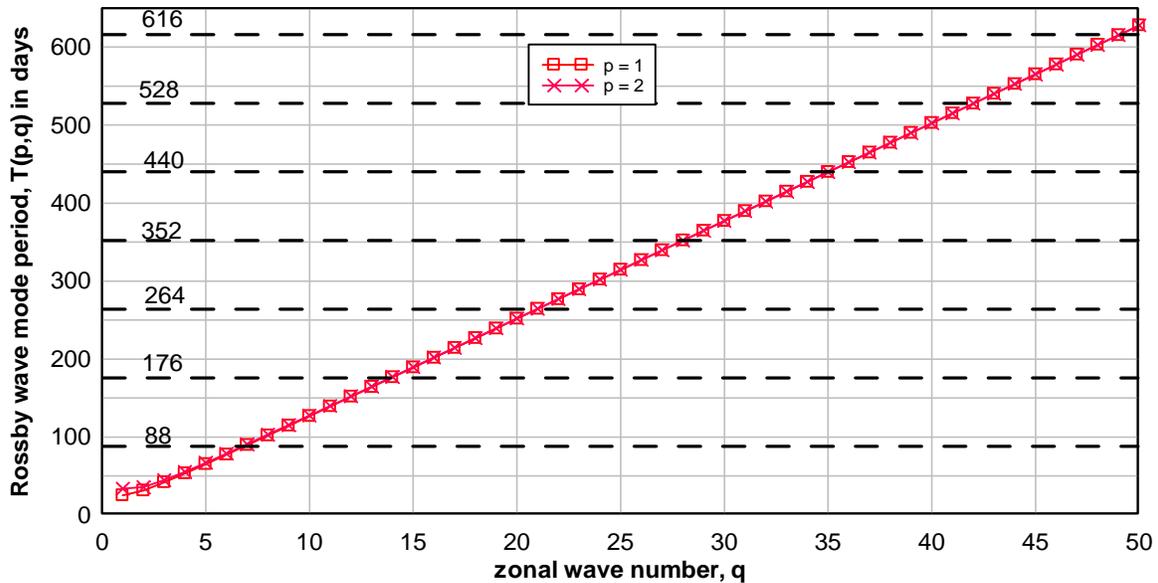

**Figure 1. The magnetic Rossby wave period from equation (1) for nodal line numbers p =1 (squares) and p = 2 (crosses) as a function of wave number q. Horizontal lines indicate the period and sub harmonic periods of Mercury.**

For modes at $q = 5$ and above there is a very small difference between the periods of the 1,q and the 2,q modes and mode periods can be found by $T(p,q) = 25.1q/2$. The reference lines in Figure 1 indicate that there is almost exact correspondence between the sub-harmonic periods of Mercury and every seventh Rossby mode period. The exact correspondence occurs because the period of Mercury, 88 days, is 3.50 times the period of photosphere rotation at the equator, 25.1 days. As a result the rotation of the equatorial photosphere is related to the orbital motion of Mercury in the sense that after two orbits Mercury, at closest approach, is above the same point on the solar surface. A Rossby wave moves around the surface of the Sun at an angular speed, relative to the surface, of $2/q^2$ times the angular speed of the suns surface, Lou et al (2003). So for higher q number modes a Rossby wave is, essentially, a stationary wave on the solar surface. As Mercury, at closest approach when the tidal effect is strongest, is above the same point every second orbit it follows that any influence of Mercury should be suited to the long term stimulation of stationary Rossby waves with mode periods close to periods associated with Mercury. Dimitropoulou et al (2008) noted that the possible Rossby wave modes are quite dense, with spacing between allowed periods of about 12 days, so that many observed Rieger type quasi-periodicities could reasonably be attributed to Rossby waves on the Sun. In their concluding remark Dimitropoulou et al (2008) noted that "an additional mechanism should be considered, which practically promotes specific modes out of the theoretically equivalent ones."

In this paper we consider the possibility that the "additional mechanism that promotes specific modes" is the tidal effect of the planet Mercury. The tidal effect is usually taken to be the height difference between high and low tide induced on the surface of the Sun by the gravitational effect of a planet. The height difference or tidal elongation due to a planet of mass M and orbital radius R at the surface of the Sun is $(3/2)(r_{Sun}^4/M_{Sun})(M/R^3)$ where $r_{Sun}$ is the radius of the Sun and $M_{Sun}$ is the mass of the Sun, Svalgaard (2011), Scafetta (2012). Thus any planetary tidal influence can be expected to vary as $1/R^3$. The orbit of Mercury is elliptical and, as the orbital radius of Mercury, $R_M$, varies by 40% from 0.47 to 0.31 astronomical units (AU) during an orbit, $1/R_M^3$ varies by 120% during each orbit of 88 days. This is the largest percentage variation of tidal effect among the planets. In the following we associate the tidal effect of Mercury with the variation of the quantity $1/R_M^3$ which varies from 9.8 $AU^{-3}$ to 34.4 $AU^{-3}$ during an orbit with average level 19.6 $AU^{-3}$.

The concept that periodic activity on the Sun may be associated with the periodicity of the planets is controversial as the calculated tidal height variations on the Sun are minute. The tidal height difference due to Mercury varies from 0.60 mm to 0.17 mm during an orbit and is therefore insignificant compared with ambient fluctuations at the solar surface or solar tacholine, (De Jager and Versteegh 2005, Callebaut et al 2012, Scafetta 2012). The history of the planetary tidal effect concept has been reviewed by Charbonneau (2002) and a revival of the planetary hypothesis has been discussed by Charbonneau (2013). The predominant idea by proponents is that tides on the Sun due to the planets somehow stimulate or trigger solar activity and the period of this resultant solar activity is similar to the period of a planet or to some more complex periodicity due to the conjunction or alignment of two or more planets with the Sun, (Hung 2007, Scafetta 2012, Abreu et al 2012, Tan and Cheng 2013). The concept of a planetary effect on solar activity remains controversial due to the lack of a convincing mechanism that would allow minute tidal surface height variations to influence magnetic activity on the Sun with research continuing e.g. Wolff and Patrone (2010), Charbonneau (2013).

There are, nevertheless, several reasons for considering a planetary effect due to Mercury.

(1) Bigg (1967), using techniques from radioastronomy for detecting periodic signals buried in noise, showed that the daily sunspot number record for the years 1850 to 1960 contained a small but consistent periodicity at the period of Mercury which is partially modulated by the positions of Venus, Earth and Jupiter.

(2) As discussed earlier and illustrated in Figure 1, there is an interesting coincidence of calculated Rossby wave mode periods with the sub harmonic periods of Mercury. For example, the ratio between the period of the 1,28 mode, 351.85718 days, and the second sub harmonic period of Mercury, 351.87704 days, is 0.99994. Thus if the "additional mechanism promoting modes" is the tidal effect of Mercury one might expect to see evidence of Rossby wave modes on the Sun at periods ~ 88 days, and at the sub harmonic periods, ~ 176 days, ~246 days, ~ 352 days, etc.

(3). There is direct observational evidence of Rossby waves on the Sun with periods close to the periodicities associated with Mercury. Sturrock and Bertello (2010) provided a power spectrum analysis of 39,000 Mount Wilson solar diameter measurements between 1968 and 1997. They fitted Rossby wave mode frequencies to the eight strongest peaks in the spectrum. The two lowest frequency peaks in their tabulation were observed at frequencies 4.04 years$^{-1}$ and 2.01 years$^{-1}$. The corresponding day periods, at 90.4 days and 182 days, are each just 3% longer than, respectively, the 88 day period of Mercury and the 176 day period of the first sub harmonic of Mercury. The periods are also very close to the periods associated with Mercury-Jupiter conjunction at 90 days and 180 days. Their observations therefore support the existence of Rossby wave modes on the Sun at periods very close to the periodicities associated with Mercury.

(4). Dominant periodicities in solar activity in the intermediate range are occasionally observed very close to the sub harmonic periods of Mercury. For example, Lean (1990) found the strongest peak in periodograms of daily sunspot area occurred at 353 days. This period is 1.003 of the third sub harmonic of the period of Mercury, 351.88 days. Chowdhury et al (2009) found prominent peaks at ~88 days and ~ 177 days in sunspot area North in solar cycle 22 close to the period and sub harmonic period of Mercury.

(5). The time variation of the tidal effect of Mercury is predictable exactly so that it is possible to establish if the time variation of the examined periodicity in solar activity is in phase with the time variation of the tidal effect. Establishing phase coherence between two variables provides a higher level test of connection over and above that obtained by comparing periods obtained by spectral analysis of the two variables.

To summarise: Mercury exerts a minute tidal force on the Sun with period 88 days, Scafetta (2012). Theoretical estimates of Rossby wave periods, Lou (2000), predict allowed modes that, we note, have periods very close to the periodicities associated with Mercury. There is direct observational evidence, Sturrock and Bertello (2010), of Rossby waves on the Sun that, we note, have periods very close to the period and the first sub harmonic period of Mercury. Rossby waves may trigger the emergence of sunspots on the surface of the Sun, (Lou et al 2003, Zaqarashvili et al 2010). However, as the modes are closely spaced and a mechanism which promotes specific modes is required, Dimitropoulou et al (2008), we consider in this paper the possible influence of Mercury.

The arrangement of the paper is as follows: Section 2 outlines the method of band pass filtering used to obtain the ~ 88 day and ~176 day period components of sunspot area. Section 3 outlines the periods associated with Mercury and Mercury-planet conjunctions and compares those periods with the periods of peaks prominent in the spectra of sunspot area. Section 4 demonstrates how the band pass filtered components of the sunspot area data are amplitude and phase modulated by episodes of sunspot area emergence. Section 5 examines the effect of episodic amplitude and phase modulation on the spectral content of sunspot area. Section 6 examines periodicity in sunspot area due to Mercury-planet conjunctions. Section 7 outlines measures of significance. Section 8 is a discussion and Section 9 a conclusion.

## 2. Methods and data.

To obtain the ~ 88 day period and ~176 day period components of sunspot area a FFT of the sunspot area data series was made. The resulting n Fourier amplitude and phase pairs, $A(f_n)$, $\phi(f_n)$, in the frequency range 0.0102 days$^{-1}$ to 0.0125 days$^{-1}$ (period range 98 to 80 days) were then used to synthesize the ~ 88 day period component of sunspot area North, denoted in this paper 88SSAN, by summing the n terms of the form $A(f_n)Cos(2\pi f_n t - \phi(f_n))$ for each day between 1876 and 2012. In the text this component is referred to as the ~88 day component. Similarly, Fourier pairs in the frequency range between 0.00483 days$^{-1}$ and 0.00653 days$^{-1}$ (period range 207 days to 153 days) were used to synthesize the ~ 176 day period component of sunspot area North denoted 176SSAN. Where data has been smoothed by, for example, a 365 day running average, the smoothed data is denoted by the suffix Snnn e.g. a 365 day running average of sunspot area North data would be denoted SSAN S365.

In the present study there are two main variables: the orbital radius of Mercury, $R_M$, and the daily sunspot area on the northern hemisphere and southern hemisphere of the Sun (SSAN and SSAS). Sunspot area is measured by the Greenwich Observatory in units of the area of one millionth of a solar hemisphere or microhems. The data is available at http://solarscience.msfc.nasa.gov/greenwch/daily_area.txt The data begins in 1874. However, due to gaps in the earliest data, we use the 50038 daily values from January 01 1876 to December 31 2012. Daily values of the orbital radii and heliocentric longitudes of the planets are available at http://omniweb.gsfc.nasa.gov/coho/helios/planet.html for 1959 to 2019. Outside this range past values were calculated by fitting a sinusoid to the 1959 to 2019 data. For example earlier values of the radius of Mercury, $R_M$, can be obtained from

$$R_M = 0.38725 - 0.07975\cos[2\pi(t - 24.5)/87.96926] \quad AU \qquad (2)$$

where time in days, t, is measured from 0 at January 01, 1995.

## 3. Periodicity in sunspot area linked to Mercury and Mercury-planet conjunctions.

Of the planets Mercury has the most elliptical orbit and has the highest angular speed around the Sun. Thus, on closest approach to the Sun there is a short, sharp increase in tidal height due to Mercury with period 88 (87.969) days. A Mercury–planet conjunction occurs when Mercury crosses the line connecting the Sun and the other planet. This results in a short, sharp increase in the tidal height due to the planet. A conjunction occurs when Mercury is on the same side of the Sun or on the opposite side so there are two conjunctions and two tidal bulges during each orbit of Mercury. Thus the period of conjunction, T, is given by the relation $1/(2T) = 1/T_1 - 1/T_2$, where $T_1$ and $T_2$ are the orbital periods of the two planets. For example, the period of Mercury-Earth conjunction, $T_{ME}$, is 58 (57.939) days. $T_{MV}$ = 72 (72.283) days and $T_{MJ}$ = 45 (44.896) days. These periods fall in the intermediate period range of solar activity as do the sub harmonics, Figure 2. Reference lines on Figure 2 indicate the average periodicity when two or more sub harmonic periods are nearly coincident.

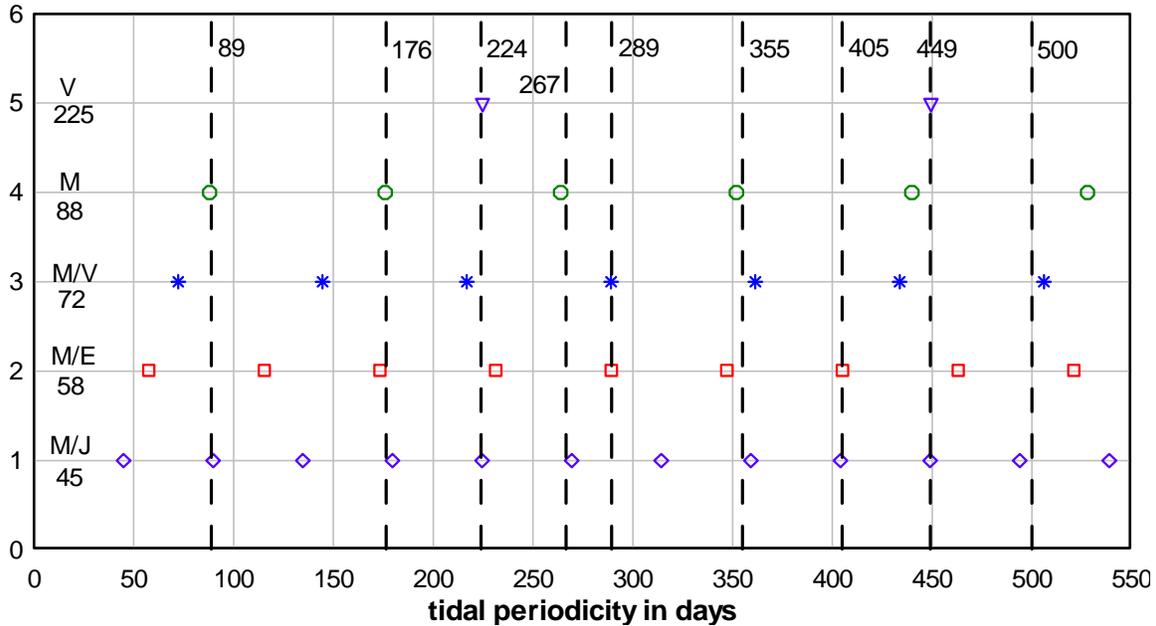

**Figure 2. Mercury, Mercury - planet periodicity and sub harmonics. Reference lines indicate the average periodicity of two or more nearly coincident periodicities.**

If the periodic emergence of sunspot area is associated with the noted planetary periodicities we would expect the spectrum of sunspot area to exhibit peaks at these periods. The spectrum of sunspot area over a solar cycle is complex and the observed period of peaks varies from solar cycle to solar cycle e.g. Chowdhury et al (2009), Lean (1990). We show later that this variation in the period of spectral peaks is due to changing episodic (amplitude and phase) modulation of sunspot emergence from solar cycle to solar cycle. Observations of this episodic modulation are described in the next section. Despite this variation in periodicity an average spectrum of sunspot area over many solar cycles should evidence planetary periodicity if there is a connection. Here we use two methods of spectral averaging, a superposition method and a standard FFT method.

The superposition method requires finding the FFT spectrum for each solar cycle, then superposing the spectra and averaging to reduce noise. Sunspot area North emergence in solar cycles 16 – 23 is 2 – 3 times higher than in solar cycles 12 – 15, see Figure 5. Therefore to increase signal to noise ratio we average superposed spectrums for cycles 16 – 23. To further increase the signal to noise ratio we ignore sunspot area contribution close to solar minima. To do this we select the time intervals over which FFT are to be averaged by drawing a 200 μhem line through the 365 day smoothed sunspot area North data in Figure 5 and average the spectra obtained for the intervals of data above this level. For example, for solar cycle 23 the data between day 44597 and day 47242 in Figure 5 provides the FFT for solar cycle 23 that forms part of the superposition average.  As the number of days included in each solar cycle is about 3000 the FFT is padded with zeroes up to 4096 to improve resolution. The average superposed FFTs for SSAN and SSAS as well as the overall average of both curves in shown in Figure 3. The resolution using this method is only moderate. Nevertheless, there is an evident consistency of peaks in the SSAN and SSAS spectra.

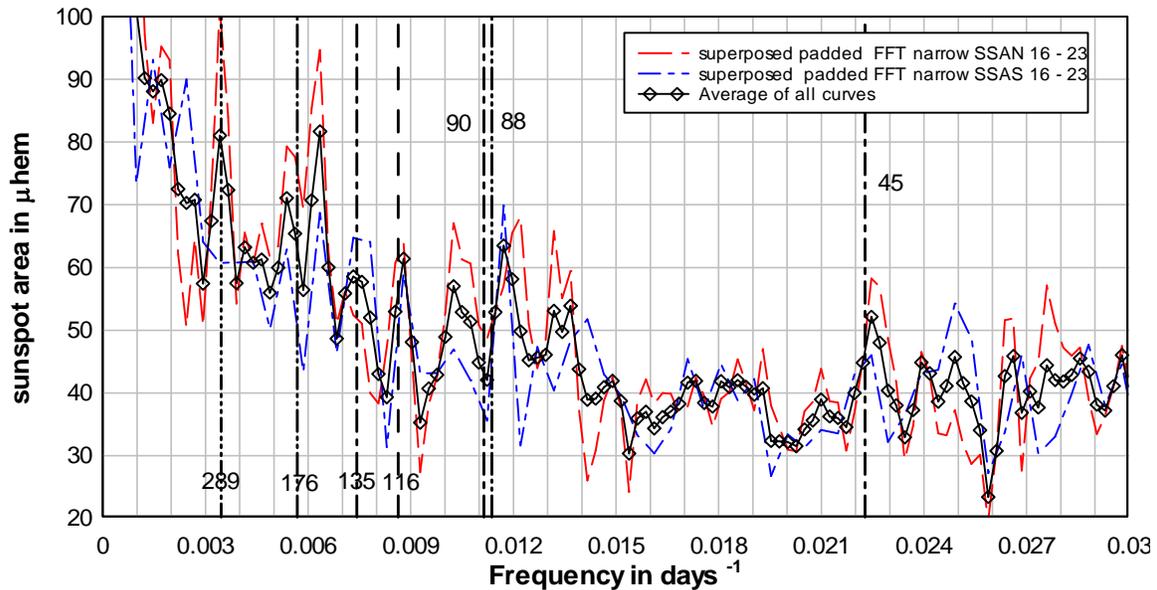

**Figure 3. The spectra of sunspot area North and sunspot area South and the average of the two spectra for solar cycles 16 – 23. The spectra were obtained by superposing the spectra for individual solar cycles and averaging as described in the text. Planetary periods in days are marked by reference lines.**

The second method involves finding a FFT spectra obtained over the entire interval covering solar cycles 16 – 23, about 30,000 days, (day 17418 to day 47650 in Figure 5). The resolution with this method is much higher and the raw SSAN and SSAS spectra comprise a dense array of peaks that is difficult to assess. To reduce this complexity the spectra are smoothed with a 20 point running average then averaged. The average spectrum, multiplied by three, is shown as the full line in Figure 4 where it is compared with the superposition average of Figure 3, broken line. Also marked on Figure 4 are reference lines indicating the main Mercury and Mercury-planet periodicities of Figure 2.

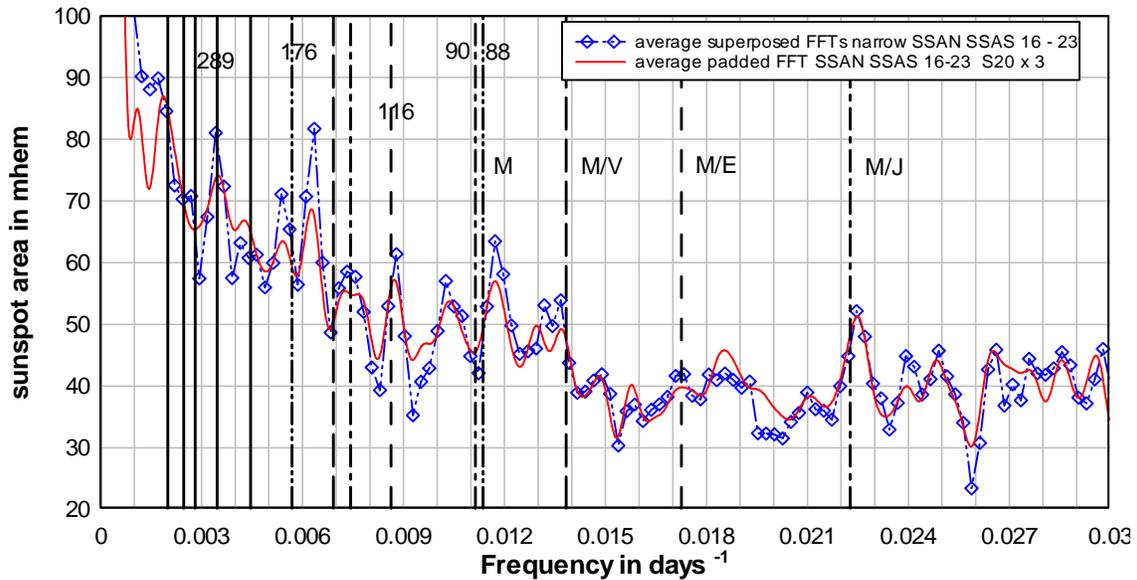

**Figure 4. Compares the average of the spectra of sunspot area North and sunspot area South for the entire interval over solar cycles 16 – 23 with the average of the superposed spectra for SSAN and SSAS as illustrated in Figure 4.   Planetary periods in days are marked by reference lines.**

While the two spectra compared in Figure 4 were obtained by quite different methods the consistency in spectral peaks is evident suggesting the observed peaks are significant. The reference lines are indicated as "dash dot" for M/J periodicity, "short dash" for M/E periodicity, "long dash" for M/V periodicity and "dash dot dot" for M periodicity. For reference periods longer than the period of the 176 day "dash dot dot" line the further reference lines, shown as full lines, refer to the averaged longer period periodicities of Figure 2.

It is evident that several of the peaks can be associated, unambiguously, with planetary periodicities, notably the 45 day M/J peak, the 116 day M/E sub harmonic peak, the 135 day M/J sub harmonic peak, and the 289 day M/E –M/V sub harmonic peak. However, there are four notable exceptions.  The 88 day and 90 day reference lines coincide with a spectrum minimum with two major peaks on either side of the minimum, at ~85 days and ~95 days.  The 176 day reference line also coincides with a spectrum minimum with two major peaks on either side of the minimum, at ~158 days and ~186 days.

The remainder of this paper is aimed at resolving the issue of these four peaks. We will show that the ~85 day and ~95 day peaks mentioned above are sideband peaks of the 88 or 90 day periodicities linked to Mercury. Similarly, the ~158 day and ~186 day peaks are sidebands peaks linked to the 176 day sub harmonic of Mercury. We will show that the sidebands can be interpreted as arising from episodic modulation (amplitude and phase) of sunspot area emergence. We note that it is difficult to distinguish, by the method of band pass filtering we use below, between Mercury and Mercury-Jupiter periodicities. Therefore the following sections focus on the Mercury 88 day and 176 day periodicities.

**4. The ~ 88 day and ~ 176 day components of sunspot area 1876 to 2012.**

The ~88 day and ~176 day components are obtained by band pass filtering as described in Section 2.

### 4.1 Episodic amplitude variation of the 88 and 176 day components.

The amplitudes of the ~88 day and ~176 day filtered components of sunspot area North are shown in Figure 5. The ~176 day component in Figure 5 is displaced by -600 microhem for clarity. Also shown, the 365 day running average of the daily SSAN and the solar cycle number.

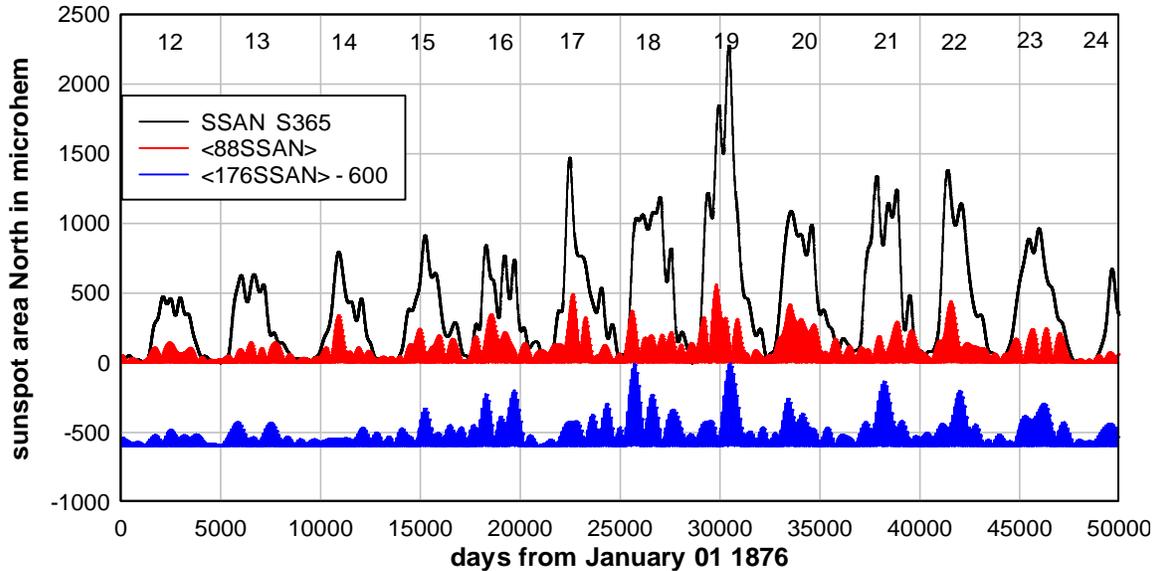

**Figure 5. The amplitude of the ~ 88 day period (red) and the ~ 176 day period (blue) components of sunspot area North from 1876 to 2012. Also shown the 365 day running average of SSAN, (black).**

It is clear from Figure 5 that variations of the ~88 day and ~176 day components are episodic. For example during solar cycle 23 there are four episodes of occurrence of the ~88 day component of SSAN, each episode about 2 years duration. The longest episodes last about 3 years e.g. the ~176 day component during cycle 19. Occasionally the episodes are discrete, i.e. not overlapping with adjacent episodes, as in cycle 23. However, more often episodes partly overlap e.g. the three episodes of the ~88 day component in cycle 20. Figure 6 shows the amplitudes of the ~88 day component and the ~ 176 day component of SSAS.

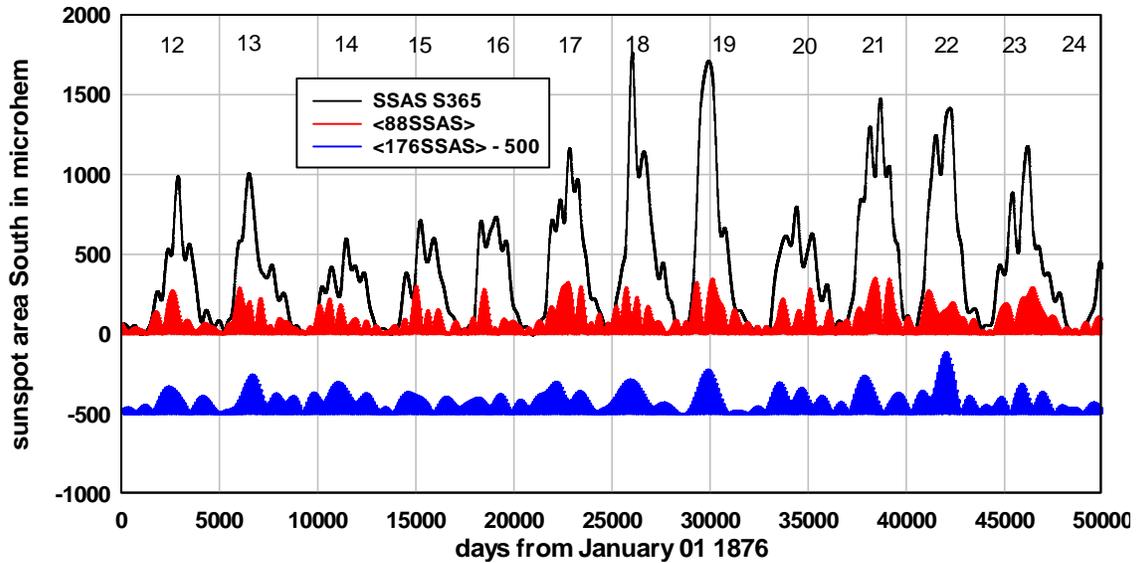

**Figure 6.** The amplitude of the ~ 88 day period (red) and the ~ 176 day period (blue) components of sunspot area South from 1876 to 2012. Also shown the 365 day running average of SSAS, (black).

**4.2 Phase of the ~88 day component variation relative to the tidal effect of Mercury.**
In addition to the amplitude modulation due to the emergence of sunspots in episodes during each solar cycle the ~88 day component is modulated in phase. A primary objective of this paper is to establish the phase of the ~88 day component of sunspot area relative to the known time variation of the tidal effect of Mercury. We can estimate the phase of the ~88 day component (y1) relative to the tidal effect (y2) by a zero crossing method or by a correlation method. The correlation method involves calculating the daily correlation using the relation y1y2/<y1y2> and smoothing the result with an 88 day running average. Either method gives essentially the same result. A correlation of 1 or 100% corresponds to the 88 day component being exactly in-phase with the tidal effect while a correlation of -1 or -100% corresponds to exactly out-of-phase. The observed phase of 88SSAN and 88SSAS relative to the tidal effect is shown in Figures 7 and 8. The figures show that the phase in successive discrete episodes usually alternates between in-phase and out-of-phase with the tidal effect. However, in some solar cycles successive discrete episodes have the same phase e.g. in solar cycle 21 three successive discrete episodes of ~88SSAN are out-of-phase with the tidal effect.

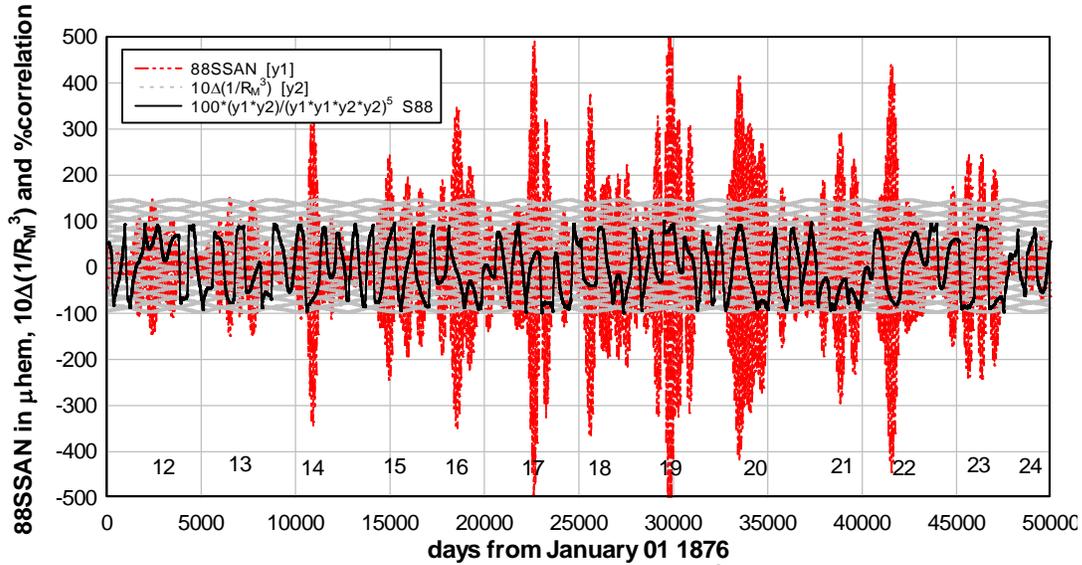
**Figure 7.** The component ~88SSAN, the tidal effect $\Delta(1/R_M^3)$ and the 88 day smoothed average correlation between the two variations expressed as a percentage.

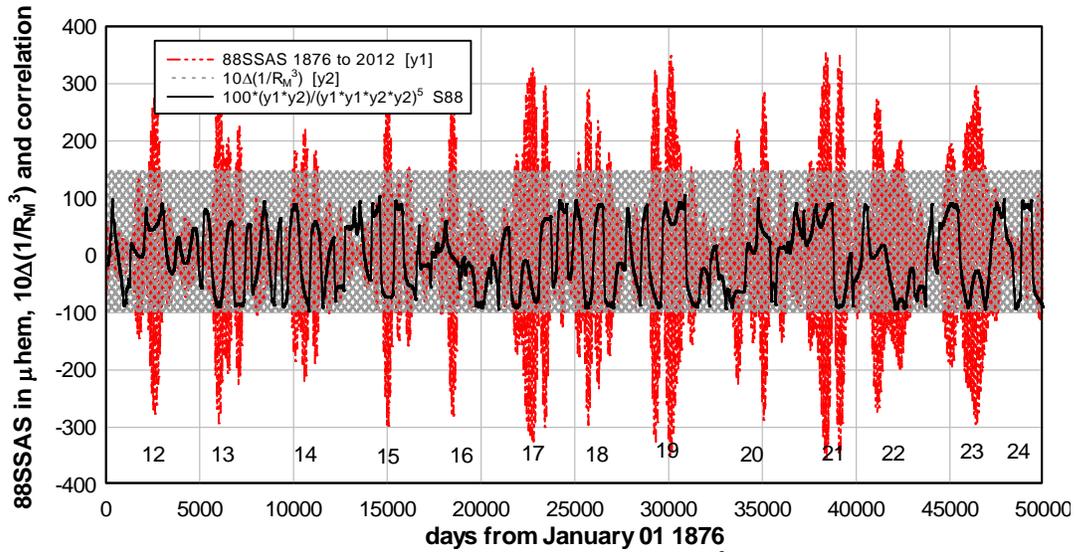
**Figure 8.** The component ~ 88SSAS, the tidal effect $\Delta(1/R_M^3)$ and the 88 day smoothed average correlation between the two variations expressed as a percentage.

Four episodes with alternating phase during a solar cycle are quite common, e.g. solar cycles 13 and 23 in the SSAN variation and solar cycles 13, 14, 15 and 18 in the SSAS variation. Two episodes of alternating phase are also common e.g. solar cycles 20 and 22 in SSAN variation and solar cycles 19 and 23 in SSAS variation. We will show in section 5 that both cases lead to sidebands at either side of a central minimum in spectra obtained over a solar cycle.

### 4.3 Phase of the ~176 day component variation relative to the tidal effect of Mercury.

The phase of the 88 day tidal effect is known precisely. However, the phase of a sub harmonic is ambiguous. The ambiguity arises in deciding which of the two peaks of the 88 day tidal effect the peak of the 176 day sub harmonic coincides with. Thus the ambiguity corresponds to a π phase shift. Here we use y1 = 150sin(2πt/175.938 + 0.4006) where t = 0 days on January 01 1876 to generate a sub harmonic which peaks at every second peak of the Mercury tidal effect. The phase relationship, estimated by the correlation method, between this function and the ~176 day components is given in Figures 9 and 10. While the actual phase relation to the tidal effect is ambiguous we note that ~176 day episodes occur with lower frequency in a solar cycle than do ~ 88 day episodes. Typically there are two episodes with alternating phase in a solar cycle.

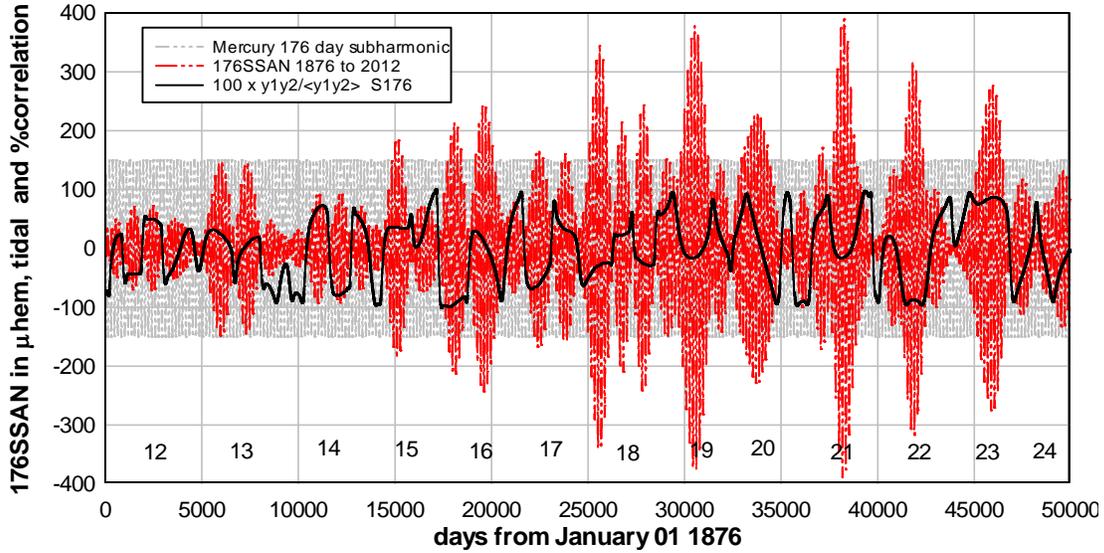

**Figure 9.** The component ~176SSAN, the 176 day variation of the first sub harmonic of the tidal effect $\Delta(1/R_M^3)$ and the 176 day smoothed average correlation between the two variations expressed as a percentage.

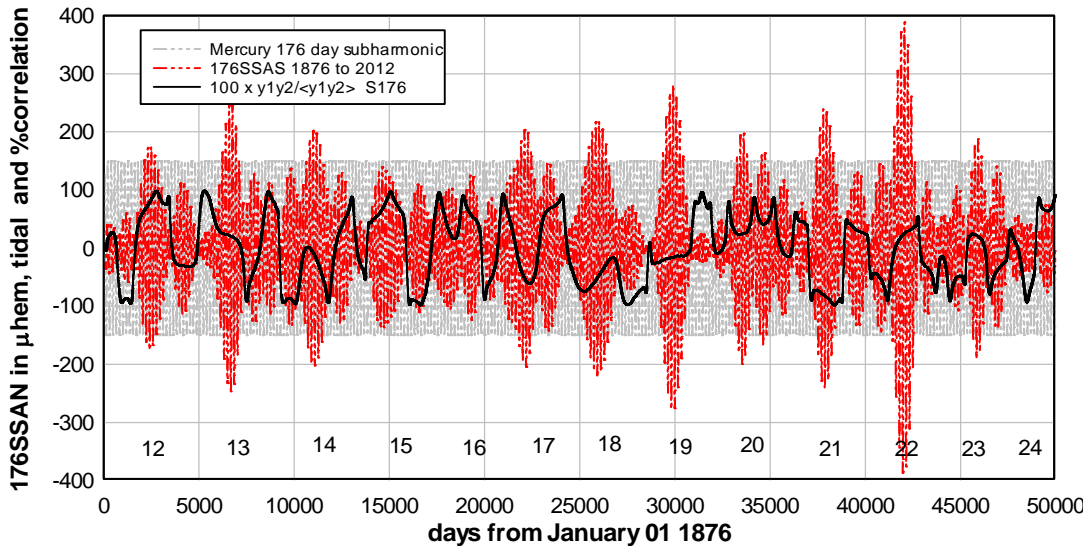

**Figure 10.** The component ~176SSAS, the 176 day variation of the first sub harmonic of the tidal effect $\Delta(1/R_M^3)$ and the 176 day smoothed average correlation between the two variations expressed as a percentage.

## 5. Spectral effect of episodic amplitude and phase modulation.

## 5.1 Simulation of the spectra of episode modulated signals.

Consider the displacement, y, associated with amplitude modulation of a sinusoid of period $T_1$.

$y = [A+\sin(2\pi t/T_m)]\sin(2\pi t/T_1)$

$\quad = [A+\sin(2\pi f_m t)]\sin(2\pi f_1 t)$

$\quad = A\sin(2\pi f_1 t) - 0.5\{\cos(2\pi(f_1+f_m)t)+\cos(2\pi(f_1-f_m)t)\}$  (3)

The terms in square brackets represent a $T_m$ day period variation that amplitude modulates the $T_1$ day period sinusoid of frequency $f_1 = 1/T_1$. Examples of the amplitude modulated variations arising from equation 3 are shown in Figures 11 and 12 with $T_1 = 88$ days and $T_m = 730$ days. It is clear that the amplitude modulation results in episodes of large amplitude variation similar to the ~88 day variation observed in the discrete episodes discussed in the previous section and observed in Figures 5 and 6. The modulation period, $T_m = 1/f_m$ in equation 3, is the time interval between the occurrence of episodes having variations with the same phase. When $A \sim 1$ there is strong amplitude modulation and side bands appear at $f_S = f_1$ +/- $f_m$ in the spectrogram, see Figure 11B. When $A \sim 0$ the modulation is said to "cross zero", the sign of the $T_1$ day sinusoid is reversed when the modulating term becomes negative, as a result there is a $\pi$ phase shift in the $T_1$ day signal between one episode and the next, and nearly all signal power is shifted from the central peak at $f_1$ to the two sideband peaks at $f_1$ +/- $f_m$, Figures 12A and 12B. The time axis of the figures is divided into 88 day intervals so that the phase of the variation in successive episodes can be followed. The sidebands for this example occur at 78.53 day and 100.06 day periods.

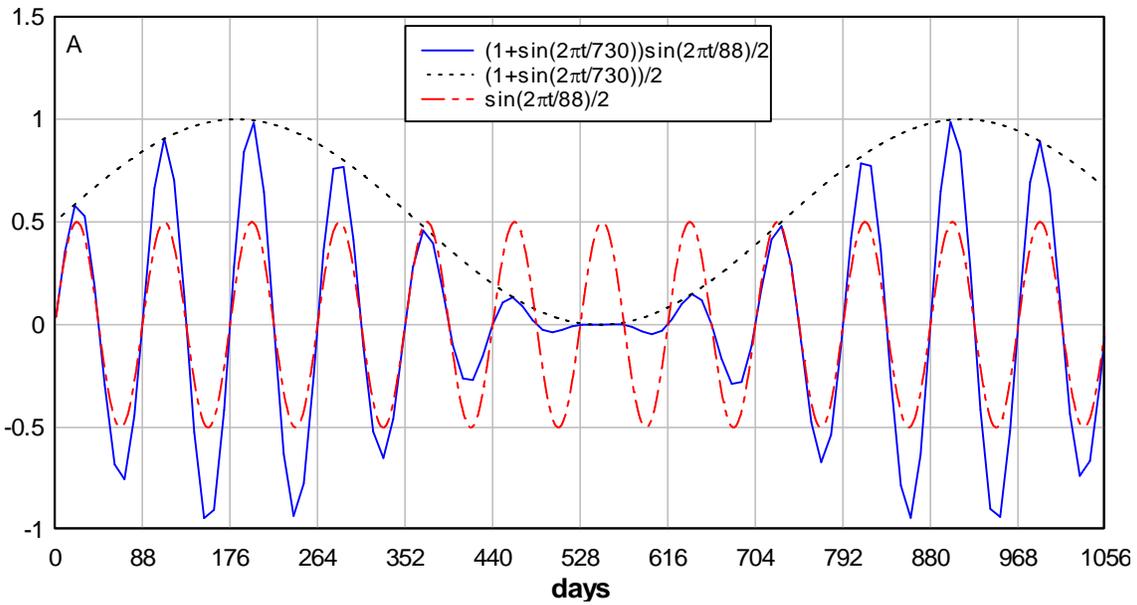

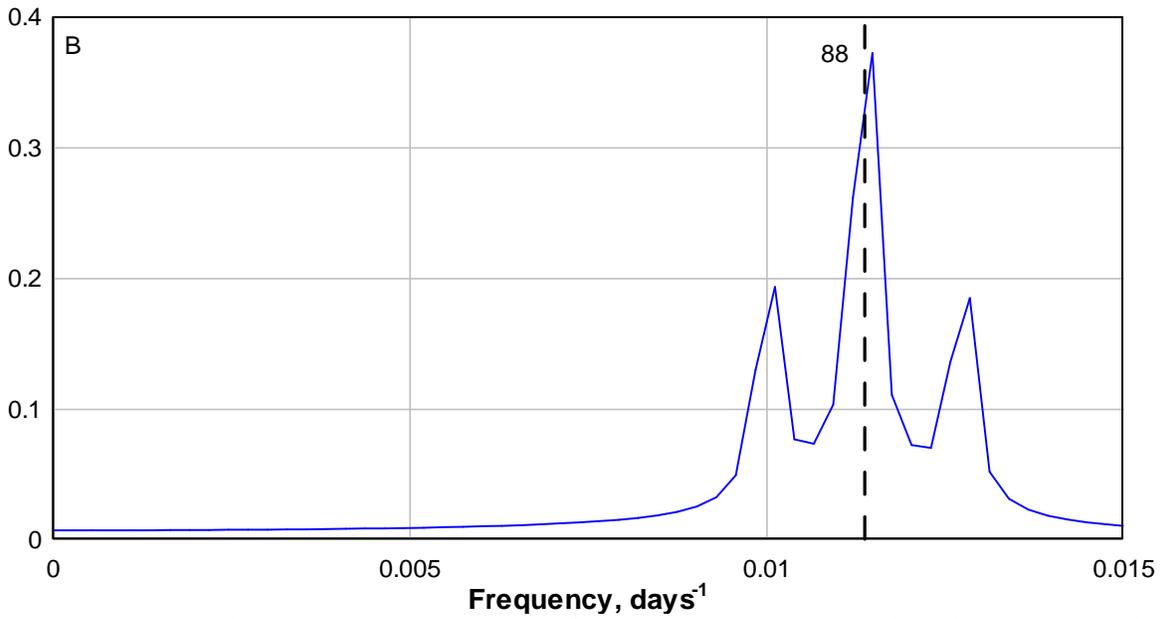

**Figure 11. The variation and the spectrogram of an amplitude modulated sinusoid when A = 1 in equation 3. The time axis of the variation is in intervals of 88 days so that any phase change can be followed. In this case there is no phase change from one modulation maximum to the next.**

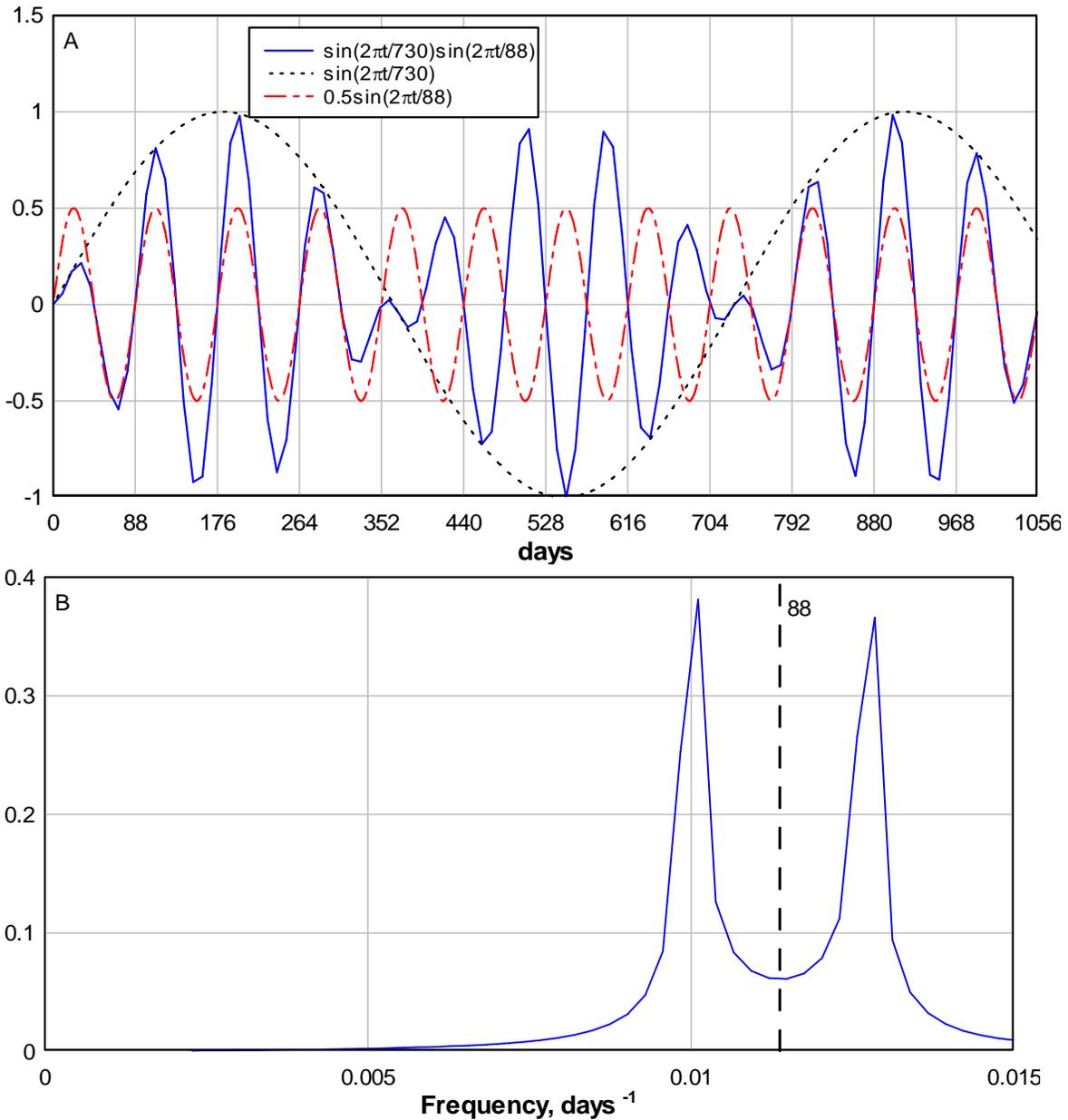

**Figure 12.** The variation and the spectrogram of an amplitude modulated sinusoid when A = 0 in equation 3. The time axis of the variation is in intervals of 88 days so that any phase change can be followed. In this case there is a π phase change from one modulation maximum to the next.

It is evident that if the variation included just one episode the spectrogram made over the variation would have a single broad peak at 88 day period. If A ~1 and the variation includes two or more discrete episodes the spectrogram would have a broad peak at 88 days period and two sidebands as in Figure 11B. If A ~ 0 and the variation includes two or more discrete episodes the spectrogram may show no peak (or only a weak peak) at 88 days and most spectral power will shift into the sidebands at 100 days and 78 days as in Figure 12B. This raises the question of distinguishing between two alternative ways of producing a time variation similar to Figure 12A and a frequency spectrum similar to Figure 12B. In one alternative the variation and the spectrum are the result of the interference of two independent, equal amplitude signals, one of period $T_x$ days and the other of period $T_y$ days. The second alternative is that the variation and the spectrum are the result of amplitude/phase modulation of a signal of period $T_1$. The necessary condition for the time variation and the spectrum to be the same in both cases is that the corresponding frequencies obey the relation $(f_x + f_y)/2 = f_1$, i.e. $f_x$ and $f_y$ are equally spaced on either side of $f_1$. In terms of periods, $1/T_x + 1/T_y = 2/T_1$ or $2T_xT_y/(T_x+T_y) = T_1$. If these relations do not hold an alternating in-phase, out-of-phase variation with the signal of period $T_1$ as illustrated in Figure 12A will not occur. In the present work the time variation of planetary tidal variation is known precisely. Therefore it is possible to compare the calculated planetary time variation, e.g. the 88 day tidal effect $\Delta(1/R_M^3)$, with the observed band pass filtered variation of sunspot area, e.g. the ~ 88SSAN component. If, as illustrated in Figure 12A discrete, successive, episodes of the ~88SSAN component alternate between being in-phase and out-of-phase with the tidal effect then the two peaks observed in the sunspot area spectrum are linked to the tidal effect periodicity via the expressions mentioned above. As shown below, an alternating $\pi$ phase change relative to the tidal effect in successive, discrete, episodes of sunspot area emergence is a useful means of identifying a planetary influence.

**5.2 Interpreting the observed spectra of sunspot area in different solar cycles.**
The simple modulation relation of equation 3 is a useful means of interpreting the effect that the episodes of sunspot area emergence during a solar cycle have on the spectra observed for that solar cycle. As an example we consider solar cycle 23. In the ~88 day component of SSAN data for solar cycle 23, Figure 13, four discrete episodes of sunspot area occur and there is a $\pi$ phase shift between the variations in succeeding episodes. As a result the spectrum can be interpreted by equation 3 with A ~ 0. However, we can also note that the phase of ~88SSAN component alternates between in-phase with the tidal effect to out-of-phase with the tidal effect in successive episodes a feature which makes it possible to distinguish between interference of two independent signals and amplitude/phase modulation of a signal at the same period as the tidal effect.

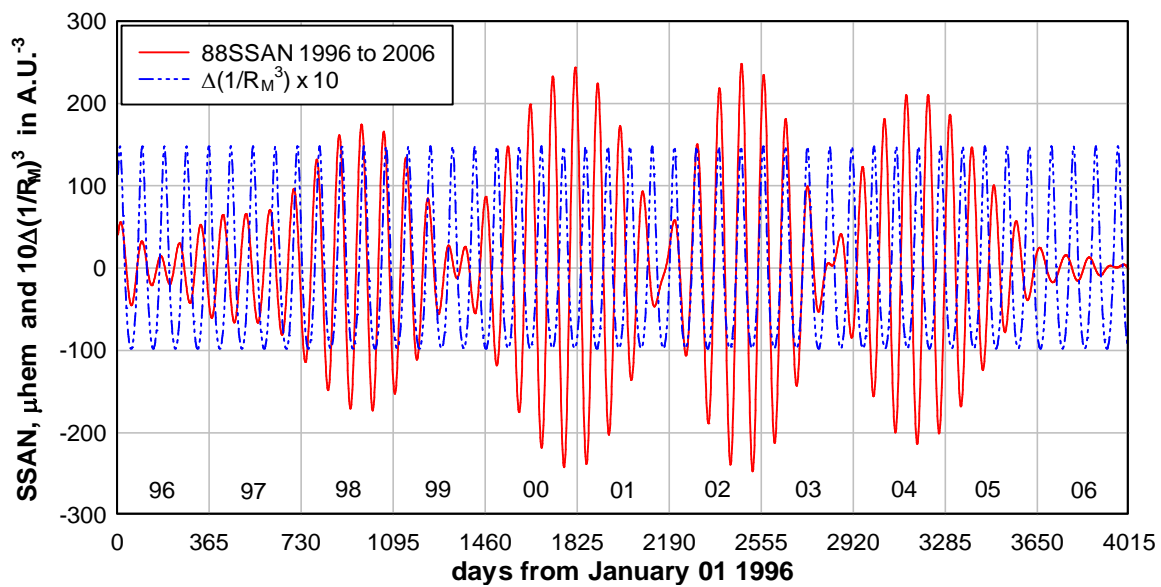

**Figure 13. The ~ 88 day component of sunspot area North for 1996 to 2006 in solar cycle 23 (full line) compared with the variation of the tidal effect of Mercury, $\Delta(1/R_M^3)$. One weak and four strong episodes of the ~88 day component of SSAN are evident with the four strong discrete episodes showing exact in-phase or exact anti-phase coherence with $\Delta(1/R_M^3)$.**

The episodes are separated, on average, by ~ 2 years or ~ 730 days. Thus the modulation period, the time interval between one episode and the next episode of the same phase, is $T_m$ = 1460 days. Thus, A = 0, $T_1$ = 88 days, and $T_m$ = 1460 days in equation 3 would simulate this case. The expected spectrogram should evidence a minimum at $T_1$ = 88 days or $f_1$ = 0.0114 days$^{-1}$ and strong sidebands at $f_1$ -/+ $f_m$ i.e. at frequencies 0.01068 days$^{-1}$ and 0.01204 days$^{-1}$. The corresponding sideband periods are 93.6 days and 83.0 days. The observed spectrogram of the unfiltered SSAN data for the eleven years between 1996 and 2006 of solar cycle 23 is shown as the blue dotted line in Figure 14. The spectrogram of the ~ 88 day band pass filtered component is also shown in Figure 14. The spectrum around the 88 day period shows the two strong sidebands on either side of a weak peak at 88 day period as expected when the ~88 day period variations in successive discrete episodes differ by π in phase. As indicated above this situation is expected for A ~ 0 in equation 3. The superposition of the raw data spectrum and the filtered data spectrum shows that the band pass filtering that produced the ~ 88 day component graphed in Figure 6 and the data in Figure 13 was wide enough to include the two relevant sidebands in cycle 23. As mentioned above a sequence of four discrete episodes during a solar cycle similar to the sequence in solar cycle 23 occurs quite often. The phase diagrams of Figure 7 and 8 show that ~88SSAN in solar cycle 13 and ~88SSAS in solar cycles 13, 14, 15 and 18 show a similar alternating phase pattern as ~88SSAN in solar cycle 23. By performing a FFT over the approximately 4000 days of each of the solar cycles it is possible to obtain similar spectra to that shown in Figure 14. The observed sideband periods in days for the two SSAN solar cycles are: sc13: 98.6, 81.9; sc23: 93.6, 83.0. For the four SSAS solar cycles the observed sideband periods in days are: sc13: 93.8, 81.9; sc14: 94.3, 81.1; sc15: 94.9, 81.8; sc18: 95.5, 81.8. The average periods are 95.1 +/- 1.8 days and 81.9 +/- 0.6 days. The average median frequency is 0.5(1/95.1 +1/81.9) = 0.01136 days$^{-1}$ corresponding to a period 88 +/- 1 days.

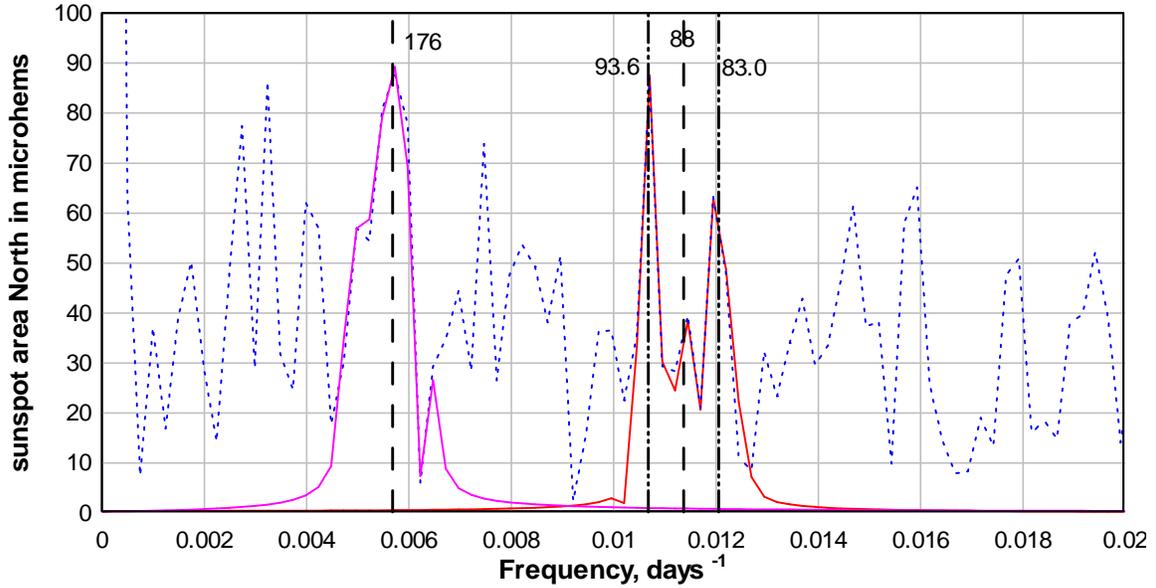

**Figure 14.** The spectrogram of daily SSAN data during the interval 1996 to 2006 in solar cycle 23, (blue dots). The spectrograms of the ~ 88 day and ~ 176 day filtered components of SSAN are superimposed for comparison.

The spectrum in Figure 14 also shows a strong peak at 176 days. From the simulations outlined above this would be characteristic of relatively weak modulation of a 176 day sinusoid, i.e. $A > 1$ in equation 3. To check this we reproduce the ~ 176 day period variation observed during cycle 23 and compare it with the tidal effect in Figure 15.

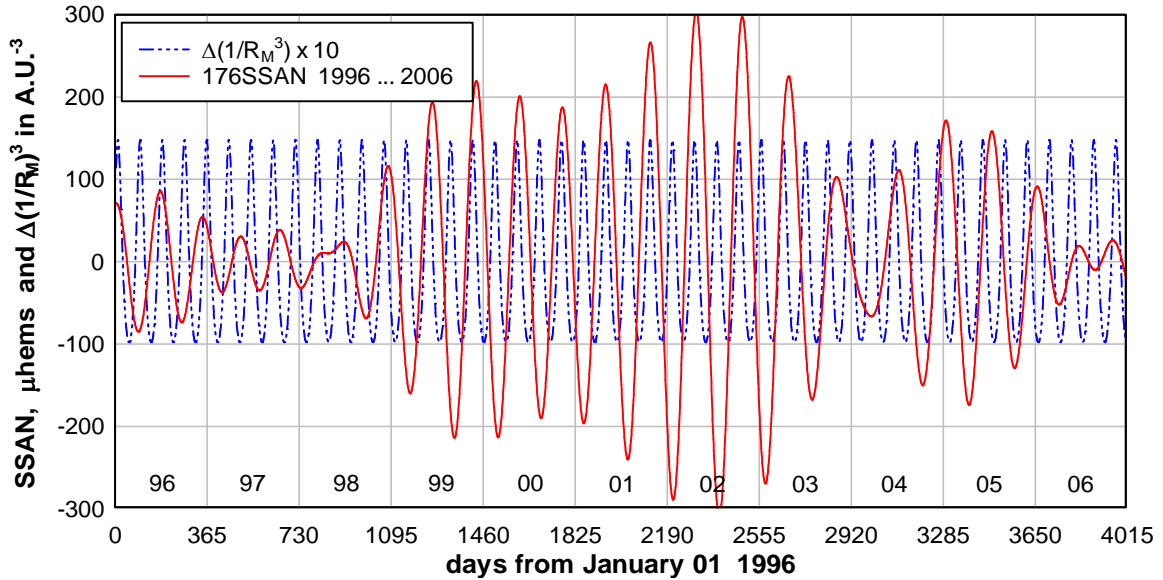

**Figure 15.** The ~176 day component of SSAN for 1996 to 2006 in solar cycle 23 (full line) compared with the variation of $\Delta(1/R_M^3)$. There is one dominant episode of ~176 day sunspot emergence during this interval with each peak in the variation in-phase with every second peak of $\Delta(1/R_M^3)$.

Figure 15 shows there is a strong episode of the ~176 day component of sunspot area North in solar cycle 23 that extends from 1999 to 2003. The variation is nearly in-phase with every second peak of the tidal effect during this interval, compare with Figure 9 which shows the phase relationship for all solar cycles. However, the variation in amplitude during this episode suggests this may be a case of two overlapping, in-phase episodes. This strong 1999 to 2003 variation overlaps a weak episode in 2004 2005. Thus the ~176 day variation of SSAN in cycle 23 is dominated by the effect of the strong 1999 – 2003 episode and a spectrogram obtained over cycle 23 should approximate the type of spectrogram obtained over a single episode i.e. the spectrogram should exhibit a strong, broad, peak at 176 days with relatively weak sidebands, as observed. Thus three major peaks in the spectrogram of Figure 14 are consistent with the type of episodes observed in cycle 23 for the ~176 day component and the ~88 day component of the SSAN data. We note that Chowdhury et al (2009) observed the same spectral peaks in SSAN data in cycle 23. We also note that, without knowledge of the occurrence of episodes of ~88 day variation in sunspot emergence, the evidence of the in-phase or and anti-phase coherence of the variations within each episode with $\Delta(1/R_M^3)$, and interpretation with equation 3, it would be difficult to discern any influence of an effect of Mercury in the spectrum because the peak at 88 days in Figure 14 is insignificant.

An example of a solar cycle that is dominated by one strong episode of the ~176 day component in SSAN and one strong episode of the ~88 day component in SSAN is, from Figure 5, solar cycle 22. We would, therefore, expect a spectrogram of sunspot area North during cycle 22 to show a strong peak at ~88 day period and a strong peak at ~176 day period and relatively weak sidebands. Using the standard FFT method we find that this is the case although we do not reproduce the spectrum here. However, in support of this interpretation, we note that Oliver and Ballester (1995), using the more powerful spectral analysis methods of Lomb-Scargle periodogram and the Maximum Entropy Method, found the most significant peak in sunspot area spectrum during cycle 22 was at 133 nHz (87 days) and noted a significant peak at 66 nHz (175 days). Similarly Chowdhury et al (2009) found significant peaks in Lombe-Scargle periodograms of sunspot area North at ~ 88 days and ~ 177 days during cycle 22.

It is clear from the above simulations and observations that sidebands of a modulated signal may provide significant sideband peaks in a spectrogram when a peak at the planetary period is small or absent. The best known Rieger periodicity is at ~160 days, Rieger (1984), Chowdhury et al (2009). Based on the analysis above we have demonstrated that this quasi-periodicity is related to the 176 day sub harmonic period of Mercury. Consider a further demonstrative example. If one wishes to discover ~160 day "quasi-periodicities" in sunspot area data one can note that the sideband frequency corresponding to 160 day period is $f_S = 1/160 = 0.00625$ days$^{-1}$. Use of the sideband relation $f_S = f_2 +/- f_m$ with $f_2 = 1/176 = 0.00568$ days$^{-1}$ indicates that the required modulation frequency is $f_m = 0.00057$ days$^{-1}$ or modulation period $T_m = 4.8$ years. Thus solar cycles with two or more discrete episodes of the ~176 day component where the successive episodes are separated by about 2.4 years and differ in phase by π radians will be solar cycles where the ~160 day "quasi-periodicity" is likely to be discovered. From Figure 5 we can see that solar cycle 18 is a good example to illustrate this effect as it has three discrete episodes of the ~176 day variation each separated by about 2.4 years. We compare the observed spectra of SSAN during solar cycle 18 with the spectrum of a simulated variation based on equation 3 with $T_2 = 176$ days and $T_m = 4.8$ years = 1754 days in Figure 16.

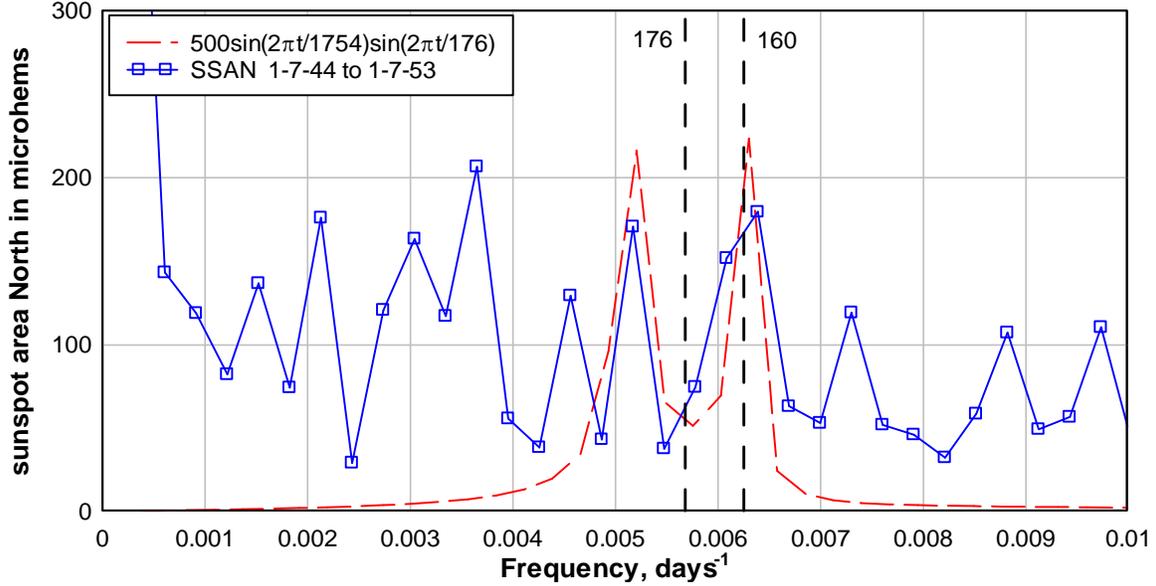

**Figure 16.** Illustrates the result of the method for "discovering" ~ 160 day periodicities in sunspot area data. The broken red line shows the spectrogram due to the variation of equation 3 with $A = 0$, $f_m = 0.00057$ days$^{-1}$ and $f_2 = 1/176$ days$^{-1}$ and the line with blue squares show the observed spectrogram of SSAN data between 1944 and 1953 in solar cycle 18.

In Figure 16 the broken red line shows the spectrogram of a simulated variation obtained from equation 3 when applying the "discovery" procedure outlined above. The line with blue squares is the actual spectrogram of the SSAN data observed during solar cycle 18, in the interval July 1 1944 to July 1 1953. Evidently a ~160 day quasi-periodicity has been "discovered" in SSAN data during this interval. There is also a longer period sideband peak at 196 days. However, according to the present analysis, the actual driving periodicity is the 176 day sub harmonic of the period of Mercury. We note again that the period of a sideband peak depends on three parameters; the constant orbital period of Mercury or periods of its sub harmonics, the varying number of episodes in the solar cycle considered and the average time interval between episodes of the same phase.

**5.3 Spectral displacement of a planetary peak due to overlapping episodes.**
So far we have considered discrete episodes in which the ~88 day component is either in -phase or out-of-phase with the tidal effect. When the episodes are overlapping, for example ~ 88SSAN during solar cycle 20, there is a progressive phase lag or phase advance in the time interval of overlap provided the variation in first episode has a different phase relation to the tidal effect than the variation in the overlapping episodes. A progressive phase lag or phase advance leads to shift of spectral power from the fundamental periodicity into a sideband periodicity at higher or lower period than the fundamental period. The effect during the approximately 4000 days of one solar cycle can be illustrated by the relation $z = \sin(2\pi t/T + \phi t/4000)$ where T is the fundamental periodicity and $\phi$ is the accumulated phase shift over the solar cycle. For example, if the fundamental periodicity T = 176 days and the total phase shift $\phi = 2\pi$ radians the relation becomes z = sin(2πt(1/176 + 1/4000)) = sin(2πt/168) and the spectral power has shifted to a peak at 168 days period.   Overlapping episodes of sunspot emergence in one hemisphere are difficult to separate. However, single discrete episodes of sunspot area emergence occasionally occur in both the Northern and the Southern hemispheres during a solar cycle. Provided the discrete episodes overlap in time we can simply add the components from both hemispheres to illustrate the effect. Solar cycle 19 is a good example as it has a dominant ~ 176SSAN episode that overlaps the single ~176SSAS episode, see Figures 5 and 6 and Figures 9 and 10. The components during solar cycle 19 are reproduced in Figure 17 as is the sum of the two components, representing the ~ 176 day component of total sunspot area. The time axis is divided into 176 day intervals so we can note that, although the variations in the discrete episodes of ~176SSAN and ~176SSAS do not exhibit a progressive phase shift the sum of the two variations does exhibit a progressive phase shift amounting to ~ 2π radians over ~ 3000 days in the solar cycle. Thus we expect the component power in total sunspot area to be shifted from 176 days to ~ 165 days. Figure 18 compares the spectra of the components ~176SSAN, ~176SSAS and the spectrum of the total during solar cycle 19.

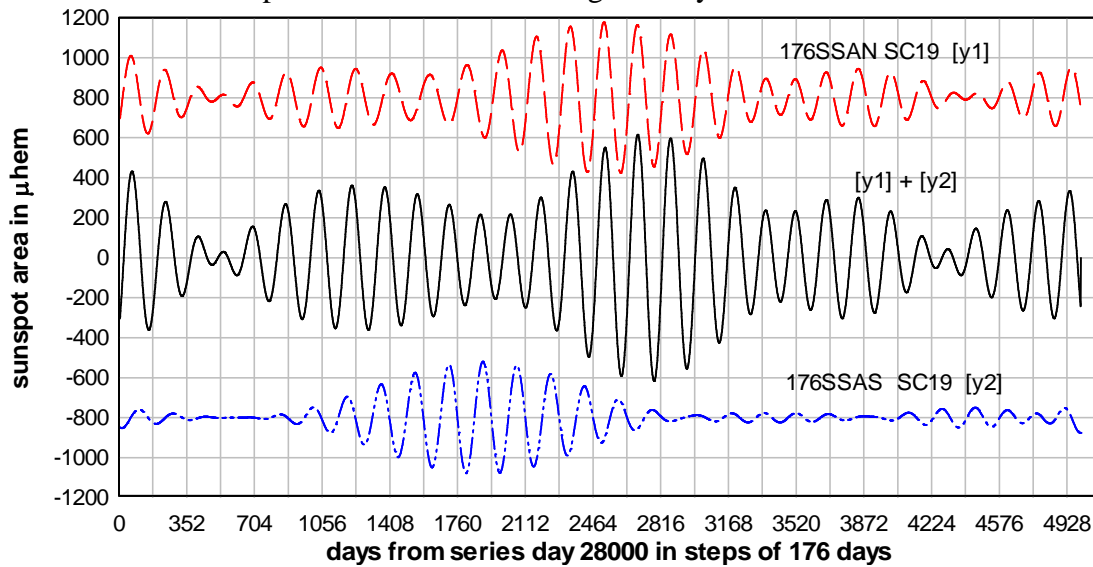

**Figure 17. The component ~176SSAN, the component ~176SSAS and their sum for solar cycle 19. The time axis is divided into 176 day intervals so that the phase shift of the components can be followed.**

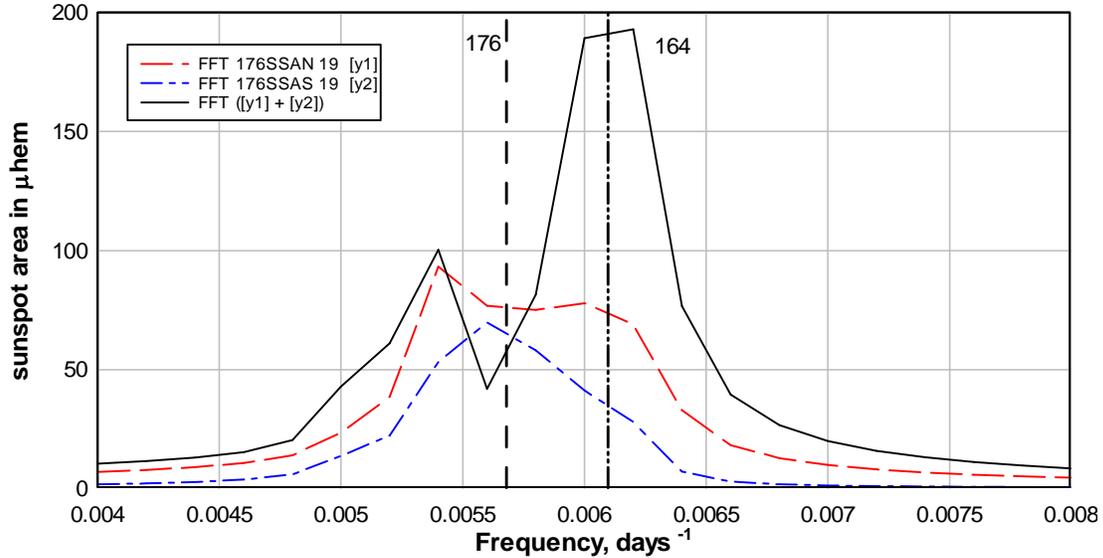

**Figure 18. The spectra of the ~176SSAN and ~176SSAS and the spectrum of the sum.**

We note that periodicity around 160 days corresponds to the Rieger periodicity that so much prior research in this area has been focused on, for example, Rieger (1984), Lean (1990), Gurgenashvili et al (2016). However, the analysis above indicates that this periodicity is derivative and not fundamental. The value of the periodicity depends on the planetary periodicity, 176 days, the duration and strength of the two planetary components in SSAN and SSAS and the extent of the overlap in time. Examination of the episodes of ~176SSAN and ~176SSAS variation illustrated in Section 3 indicate that episodes in the two hemispheres will frequently overlap during solar cycles. It follows that the spectra of total sunspot area (SSAT) during various solar cycles will often exhibit periodicities that have been shifted down from 176 days towards 160 days or up from 176 days towards 190 days. We have not directly examined the periodicity in total SSAT in this paper. However, Gurgenashvili et al (2016) have specifically examined this Rieger type periodicity in SSAT, in the period range 140 – 200 days, for solar cycles 14 to 24. Their observations, their Figure 1, are consistent with the above analysis i.e. that the spectral peak of SSAT in this period range is either shifted up towards 190 days or down towards 160 days or two spectral peaks occur on either side of 176 days at ~ 160 days and ~190 days. We note that Gurgenashvili et al (2016) provide a different explanation of the data to the explanation presented above.

**6. Periodicity in sunspot area due to Mercury – planet conjunctions.**

In the average spectra of sunspot area in Figures 3 and 4 it is evident that spectral splitting, i.e. the presence of two sidebands on either side of a planetary period, is observed for the 88 day and 176 day Mercury periodicity but not for the 45 day M/J, the 116 day M/E sub harmonic, the 135 day M/J sub harmonic or the 289 day M/V – M/E sub harmonic periodicity all of which appear to be single, broad spectral peaks centred on the corresponding Mercury – planet periodicities. Here we examine why these planetary periodicities are associated, in average spectra, with single peaks.

## 6.1 The 116 day Mercury – Earth sub harmonic periodicity.

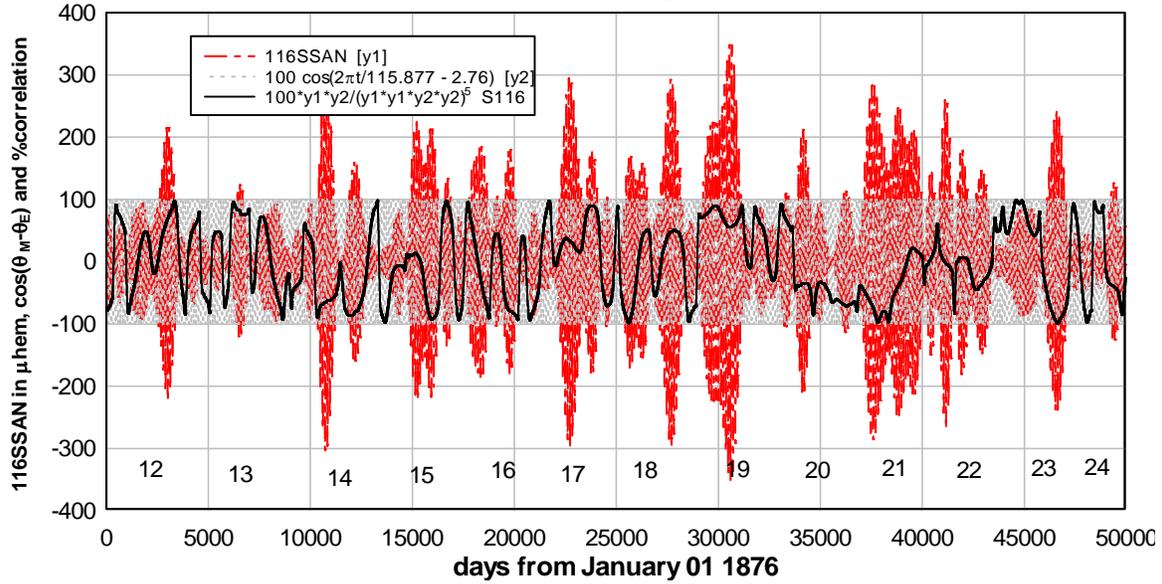

**Figure 19.** The component ~116SSAN, the 116 day variation of the first sub harmonic of the Mercury – Earth conjunction and the 116 day smoothed average correlation between the two variations expressed as a percentage.

The 116 day periodicity corresponds to the first sub harmonic period of the 58 day M/E conjunction period. The time variation of the variation in tidal effect due to the 58 day conjunction can be found from the relation $I_{ME} = <\cos(\theta_M - \theta_E)>$ where $\theta_M$ and $\theta_E$ are daily values of the heliocentric longitudes of Mercury and Earth respectively and $I_{ME}$ is the alignment index, Hung (2007). A maximum in tidal bulge due to this conjunction occurs when $I_{ME} = 1$ and with period 58 days. The 116 day first M/E sub harmonic varies as $\cos(\theta_M - \theta_E)$ i.e. there is a peak at every second peak of $I_{ME}$. However, there is a $\pi$ phase ambiguity arising from which second peak of $I_{ME}$ is coincident with a peak in $\cos(\theta_M - \theta_E)$. We make an arbitrary choice and use the function $100\cos(2\pi t/115.877 - 2.76)$ where t = 0 on January 01 1876 to represent the variation of $100\cos(\theta_M - \theta_E)$ in Figure 19. The correlation between this function and the ~ 116SSAN component is calculated as described earlier and shown as the full line in Figure 19. It should be noted that the $\pi$ phase ambiguity might lead to an inversion of the phase difference curve in Figure 19. For example, Figure 19 indicates that during solar cycle 19 twenty cycles of the ~116 day component vary in-phase with the first sub harmonic of the M/E conjunction. Alternatively, taking into account the ambiguity, the 20 cycles of the ~ 116 day component are varying out-of-phase with the first sub harmonic.

It is evident that during the stronger solar cycles e.g. cycles 14, 17, 19, 21, 22 and 23 the phase difference between the ~116 day SSAN component and $\cos(\theta_M - \theta_E)$ remains relatively constant as compared, for example, with the more frequent alternation of the phase relation between the tidal effect, $\Delta(1/R_M^3)$, and the ~88SSAN or ~88SSAS components, see Figures 7 and 8. As outlined in Section 5 a constant phase relationship corresponds to weak episodic modulation and results in a spectral peak at the planetary period with weak or non-existent sidebands. Therefore in average spectra over solar cycles 16 – 23 we would expect a single peak at 116 day period.

## 6.2 The 289 day Mercury – Earth, Mercury – Venus, sub harmonic periodicity.

A single strong peak at 289 days is evident in the spectra of Figures 3 and 4. According to Figure 2 this coincides with the fourth sub harmonic of the 58 day M/E conjunction period and the third sub harmonic of the 72 day M/V conjunction period. The episodic variation of the ~289 day SSAN variation is shown in Figure 20 where it is evident that there is a single episode in nearly all the solar cycles. On this basis we expect a single, strong peak at 289 days in averaged spectra of sunspot area as observed.

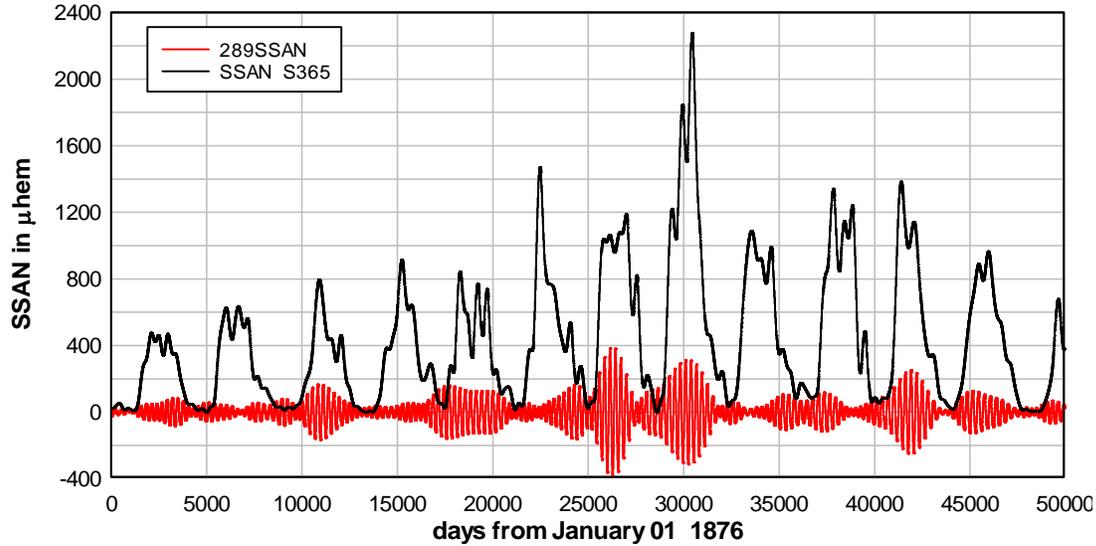

**Figure 20. Compares the variation of the component ~ 289SSAN with the variation of SSAN. It is evident there is usually only one episode of the ~ 289SSAN variation in each solar cycle.**

If the alignment indices $I_{ME}$ and $I_{MV}$ are added we find that the two variations interfere to form beats with peak beat amplitude occurring at intervals of 291.96 days, Figure 21. The occurrence of beats at sub harmonic periodicity may be the reason why emergence of sunspot area occurs periodically at sub harmonics of conjunction periods. Figure 21 compares the ~289SSAN component for solar cycle 22 with the combination of $I_{ME}$ and $I_{MV}$. Also shown is a sinusoidal variation fitted to the combination of $I_{ME}$ and $I_{MV}$, $70\sin(2\pi t/291.96 - 2.6)$ where t is measured from January 01 1876.

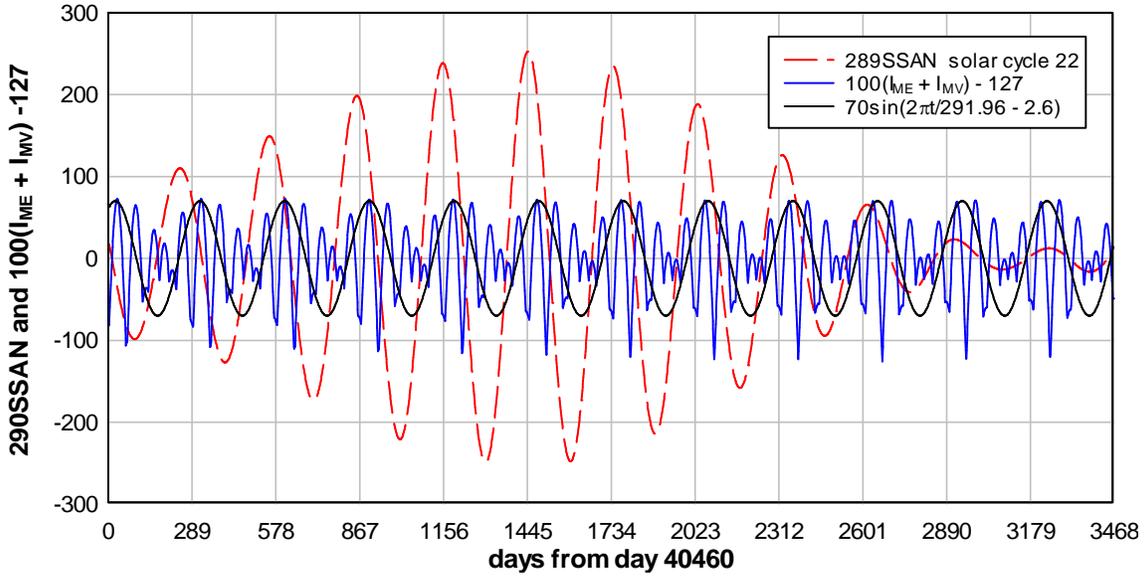

**Figure 21. Compares the ~289SSAN component during solar cycle 22 with the sum of the alignment indices $I_{ME}$ and $I_{MV}$. Also shown a fitted sinusoid that peaks at the peak of the interference pattern between $I_{ME}$ and $I_{MV}$.**

The phase difference between this sinusoidal variation and the ~289SSAN variation is shown in Figure 22. The phase difference alternates slowly so that the phase change in any solar cycle tends to be modest, leading again to the expectation of a single peak at ~ 289 days period in average spectra of SSAN and SSAS.

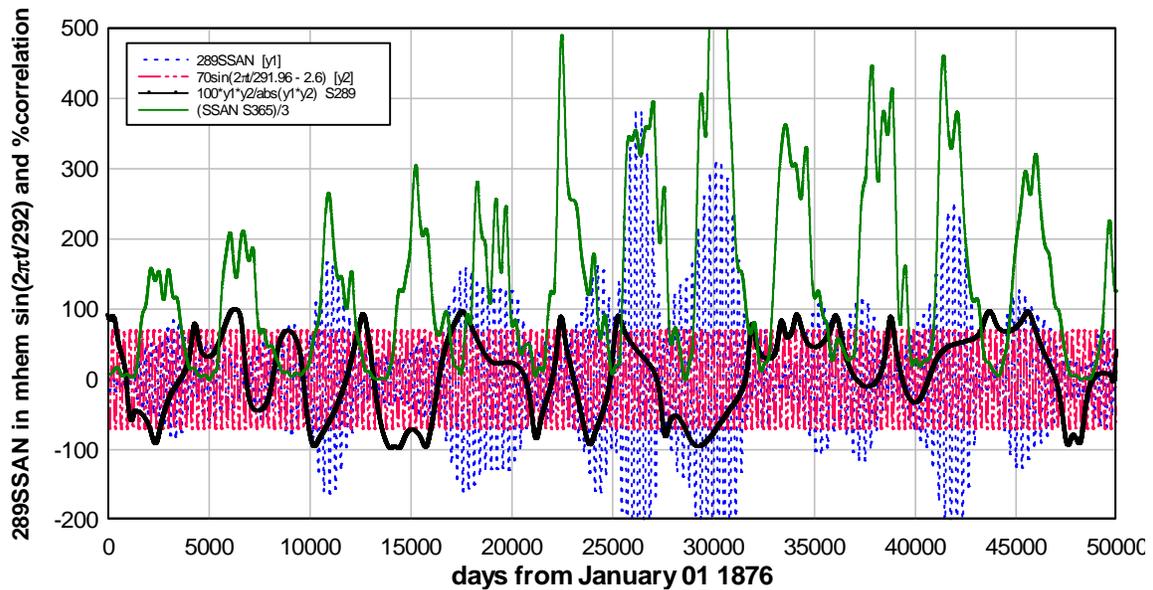

**Figure 22. Shows the percentage correlation between the ~ 289SSAN component and a sinusoid peaking at the beat maxima of the interference pattern of $I_{ME}$ and $I_{MV}$. Also shown the variation of SSAN.**

The analysis in this section provides an explanation of why, in spectra, we observe single peaks at 116 day period and 289 day period whereas 88 day and 176 day periodicity results in spectral splitting. However, it does not answer the deeper question of why the episodic emergence and phase relationship is different in these two cases.

We note that the phase in successive episodes of ~289SSAN does alternate but with long intervals between episodes with approximately the same phase. On average the interval, $T_m$, is ~10,000 days. This would result in spectral splitting of the peak at 289 days into sidebands at frequencies 0.00346 +/- 0.00010 days$^{-1}$ or periods 289 +/- 8 days. This results in complex fine detail in spectra well beyond the resolution used in the average spectra of Figures 3 and 4.

## 7. Significance

The hypothesis of this paper is that sunspot emergence is a complex systematic process influenced by many of the planetary periodicities outlined in Figure 2. It is well known that the result of a complex systematic process may be difficult to distinguish from the result of a random process or noise. This section canvases options for assessing significance rather than attempting a full statistical assessment of significance.

The primary evidence presented for planetary influence is the presence of prominent peaks that can be linked to planetary periodicities in spectra from both the sunspot area North data and sunspot area South data. In the spectra of the SSAN and SSAS data shown in Figures 3 and 4 the eight most prominent peaks have periods that are either coincident with planetary periods or can be associated with sidebands to planetary periods. A test for significance then is how often a similar set of peaks would occur in the spectra obtained from similar length series of red noise. To illustrate how this might be tested we generated 30,000 values of red noise using $z(1) = 0.9z(0) + rnd$ where rnd is white noise. Figure 23 compares the resulting noise spectrum with the average spectra of SSAN and SSAS for cycles 16 – 23, Figure 4. Obviously the set of noise spectral peaks is not similar to the set of spectral peaks due to the sunspot area data. However, in terms of significance it would be a matter of generating and analyzing many noise series until a matching set of peaks was found and the probability of this set of peaks occurring in the spectra of red noise could be determined. However, it is clear from Figure 23 that some peak coincidence will occur with any noise series spectrum. For example in Figure 23 noise peaks occur close to the 58 day M/E period, the 90 day M/J period and the 176 day Mercury sub harmonic period.

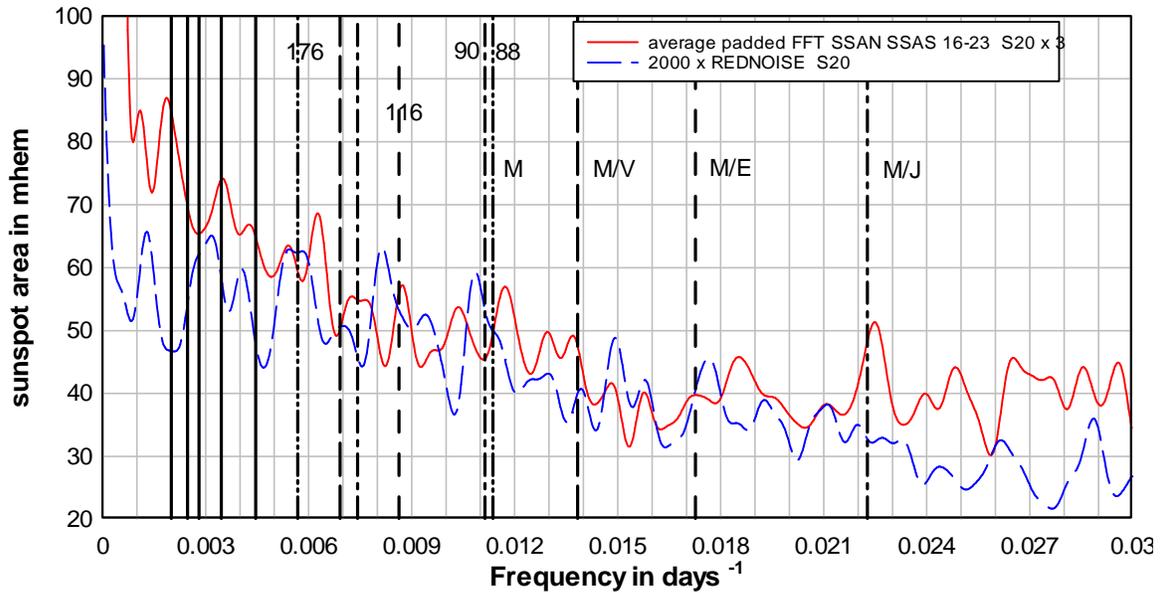

**Figure 23. Compares the averaged SSAN and SSAS spectrum for solar cycles 16-23 (full line) with the spectrum obtained from a similar length series of red noise (broken line).**

The secondary evidence of planetary influence is the distinctive alternating $\pi$ phase change relative to the tidal effect in discrete episodes of the ~ 88SSAN and ~88SSAS components of sunspot area during a solar cycle. The discussion in section 5.1 suggests that this indicates a planetary influence. However, if a red noise series is band pass filtered and the phase of the ~88 day component of red noise is compared with the 88 day period tidal effect similar alternating in-phase and out-of-phase intervals can occasionally occur as illustrated in Figure 24. For example, an interval of alternating $\pi$ phase change occurs between day 35000 and 38000 in Figure 24. Four such intervals occur in 88SSAN, see Figure 8 in the Revision. Thus the significance might be determined by generating a series of red noise phase diagrams similar to that in Figure 24 and determining the probability of four intervals of alternating $\pi$ phase change similar to those in Figure 8 occurring.

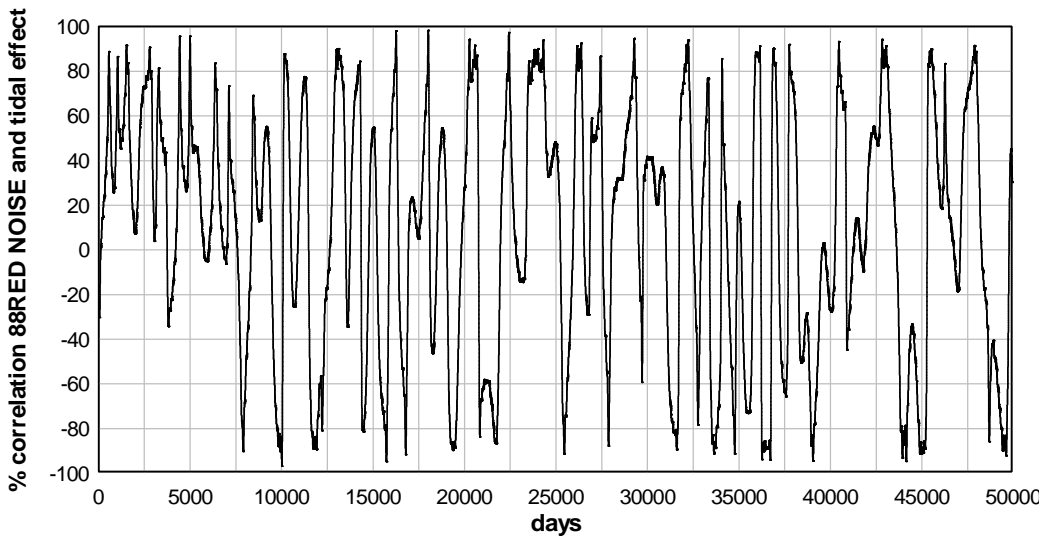

**Figure 24. The percentage correlation between the ~88 day period band pass filtered component of red noise and the 88 day period tidal effect due to Mercury.**

## 8. Discussion

There are three principal findings in this article. The first finding is that the spectra of SSAN and the spectra of SSAS are both dominated by peaks linked to planetary periodicities. In particular the M/J 45 day conjunction period, the M/E 116 day first sub harmonic conjunction period, the 135 day sub harmonic conjunction period and the 289 day M/E – M/V sub harmonic conjunction period can be linked by direct coincidence of the peaks in sunspot area spectra with 45, 116, 135 and 289 day periods. The 88 day Mercury tidal period can be indirectly, but unambiguously, linked to the sideband peaks at ~85 days and ~95 days. Similarly the Mercury 176 day first sub harmonic tidal periodicity can be indirectly linked to the sideband peaks at ~158 days and ~186 days. Thus the eight most prominent peaks in the intermediate range spectra of sunspot area can be linked to planetary periodicities associated with Mercury. There are indications that longer period spectral peaks may also be linked to planetary periodicity.

The second finding is that the ~88 day and ~176 day components of sunspot area emerge in episodes of higher amplitude during the maximum part of each solar cycle. The episodes are typically of ~2 year duration for the ~88 day components and of ~4 year duration for the ~176 day component. The third finding is that when episodes of the ~88 day component are discrete the phase of the component within successive episodes tends to alternate between in-phase and out-of-phase with the Mercury tidal effect. This provides a means of distinguishing between the observed peaks being at sideband periods of the planetary periodicity or the observed peaks arising from the interference of two sunspot area variations at periods independent of planetary periodicity. Why the phase inversion between successive discrete episodes of periodic sunspot emergence occurs is unknown. It may be related to the fact that there are two Rossby wave modes at each period or it may be related to the fact that there are two nearly coincident planetary periods in these cases, e.g. the 88 day Mercury period and the 90 day M/J sub harmonic period.

There is prior evidence that the emergence of sunspots may be related to the presence of Rossby waves on the Sun, (Lou 2000, Lou et al 2003, Dimitropolou et al 2008, Zaqarashvili et al 2010). This paper advances the idea that Rossby waves, having mode periods very close to the Mercury and Mercury – planet periods may have, over millennia, become synchronised to the tidal effect of Mercury and its sub harmonics. We noted that the rotation period of the solar surface at the equator, ~25.1 days, is such that closest approach of Mercury to the Sun occurs above the same point on the solar equator every seven solar rotations thereby favouring stimulation of stationary, equatorially trapped Rossby waves on the Sun by the tidal effect of Mercury. The calculated tidal displacement due to Mercury is of the order 1 mm, Scafetta (2012). Rossby waves with periods close to 88 days and 176 days and observable amplitude have been reported by Sturrock and Bertello (2010). The vertical displacement due to an observable Rossby wave is expected to be about 70 km, Lou (2000). Thus long term stimulation of Rossby waves by the tidal effect of Mercury, if it did occur, could provide a 70 million times amplification of the tidal effect. Rossby waves of significant amplitude at planetary periods might trigger unstable magnet flux to emerge with the same periodicity on the solar surface to form sunspots.

## 9. Conclusion

The observations presented above support a connection between the tidal effect of Mercury and sunspot emergence on the Sun. However, the ideas outlined above for the mechanism that might support a connection are very speculative. In particular concepts of: (a), how the minute periodic variations in surface height on the Sun due to the tidal effect of Mercury could excite the long term growth of Rossby waves to observable amplitude; (b), how Rossby waves trigger the emergence of magnetic flux on the Sun; (c), why episodes of emergence occur over varying intervals of a few years; and (d), why, when episodes of emergence are discrete, there is a $\pi$ phase change in the periodic variation of sunspot area from one episode to the next.

On the other hand the analysis of sunspot area data in terms of planetary influence provides a reasonably consistent basis for understanding the observable data, e.g. why planetary periodicity is prominent in the average spectra of sunspot area; why some spectral peaks are split into sidebands and others are not; why the sideband peaks vary in period from solar cycle to solar cycle; why sideband peaks are displaced up or down in period in spectra obtained over overlapping episodes; why spectral splitting occurs for some planetary periods while for other planetary periods single peaks occur; and how constructive interference of the tidal effect at two closely similar planetary periodicities might explain the occurrence of spectral peaks at sub harmonics of the planetary periodicities. One of the principal and enigmatic characteristics of intermediate range periodicity is intermittency, i.e. spectral peaks occur in some solar cycles and are absent in other solar cycles. The present analysis explains this conundrum. If a single episode of periodic sunspot emergence occurs in a solar cycle a spectral peak at the planetary period is present. If two or more discrete episodes occur during a solar cycle the planetary spectral peak splits into two sideband peaks which are variable in period depending on the detailed nature of the episodes. Alternatively, if the episodes overlap the planetary spectral peak is shifted up or down to a sideband period which is variable in period, the variation in period depending on the detailed nature of the episodes.

**Acknowledgment**

**References.**

**Figure Captions**

**Figure 1.** The magnetic Rossby wave period from equation (1) for nodal line numbers p =1 (squares) and p = 2 (crosses) as a function of wave number q. Horizontal lines indicate the period and sub harmonic periods of Mercury.

**Figure 2.** Mercury, Mercury - planet periodicity and sub harmonics. Reference lines indicate the average periodicity of two or more nearly coincident periodicities.

**Figure 3.** The spectra of sunspot area North and sunspot area South and the average of the two spectra for solar cycles 16 – 23. The spectra were obtained by superposing the spectra for individual solar cycles and averaging as described in the text. Planetary periods in days are marked by reference lines.

**Figure 4.** Compares the average of the spectra of sunspot area North and sunspot area South for the entire interval over solar cycles 16 – 23 with the average of the superposed spectra for SSAN and SSAS as illustrated in Figure 4.   Planetary periods in days are marked by reference lines.

**Figure 5.** The amplitude of the ~ 88 day period (red) and the ~ 176 day period (blue) components of sunspot area North from 1876 to 2012. Also shown the 365 day running average of SSAN, (black).

**Figure 6.** The amplitude of the ~ 88 day period (red) and the ~ 176 day period (blue) components of sunspot area South from 1876 to 2012. Also shown the 365 day running average of SSAS, (black).

**Figure 7.** The component ~88SSAN, the tidal effect $\Delta(1/R_M^3)$ and the 88 day smoothed average correlation between the two variations expressed as a percentage.

**Figure 8.** The component ~ 88SSAS, the tidal effect $\Delta(1/R_M^3)$ and the 88 day smoothed average correlation between the two variations expressed as a percentage.

**Figure 9.**  The component ~176SSAN, the 176 day variation of the first sub harmonic of the tidal effect $\Delta(1/R_M^3)$ and the 176 day smoothed average correlation between the two variations expressed as a percentage.

**Figure 10.**  The component ~176SSAS, the 176 day variation of the first sub harmonic of the tidal effect $\Delta(1/R_M^3)$ and the 176 day smoothed average correlation between the two variations expressed as a percentage.

**Figure 11.** The variation and the spectrogram of an amplitude modulated sinusoid when A = 1 in equation 3. The time axis of the variation is in intervals of 88 days so that any phase change can be followed. In this case there is no phase change from one modulation maximum to the next.

**Figure 12.** The variation and the spectrogram of an amplitude modulated sinusoid when A = 0 in equation 3. The time axis of the variation is in intervals of 88 days so that any phase change can be followed. In this case there is a $\pi$ phase change from one modulation maximum to the next.

**Figure 13.** The ~ 88 day component of sunspot area North for 1996 to 2006 in solar cycle 23 (full line) compared with the variation of the tidal effect of Mercury, $\Delta(1/R_M^3)$.  One weak and four strong episodes of the ~88 day component of SSAN are evident with the four strong discrete episodes showing exact in-phase or exact anti-phase coherence with $\Delta(1/R_M^3)$.

**Figure 14.** The spectrogram of daily SSAN data during the interval 1996 to 2006 in solar cycle 23, (blue dots). The spectrograms of the ~ 88 day and ~ 176 day filtered components of SSAN are superimposed for comparison.

**Figure 15.** The ~176 day component of SSAN for 1996 to 2006 in solar cycle 23 (full line) compared with the variation of $\Delta(1/R_M^3)$. There is one dominant episode of ~176 day sunspot emergence during this interval with each peak in the variation in-phase with every second peak of $\Delta(1/R_M^3)$.

**Figure 16.** Illustrates the result of the method for "discovering" ~ 160 day periodicities in sunspot area data. The broken red line shows the spectrogram due to the variation of equation 3 with A = 0, $f_m = 0.00057$ days$^{-1}$ and $f_2 = 1/176$ days$^{-1}$ and the line with blue squares show the observed spectrogram of SSAN data between 1944 and 1953 in solar cycle 18.

**Figure 17.** The component ~176SSAN, the component ~176SSAS and their sum for solar cycle 19. The time axis is divided into 176 day intervals so that the phase shift of the components can be followed.

**Figure 18.** The spectra of the ~176SSAN and ~176SSAS and the spectrum of the sum.

**Figure 19.** The component ~116SSAN, the 116 day variation of the first sub harmonic of the Mercury – Earth conjunction and the 116 day smoothed average correlation between the two variations expressed as a percentage.

**Figure 20.** Compares the variation of the component ~ 289SSAN with the variation of SSAN. It is evident there is usually only one episode of the ~ 289SSAN variation in each solar cycle.

**Figure 21.** Compares the ~289SSAN component during solar cycle 22 with the sum of the alignment indices $I_{ME}$ and $I_{MV}$. Also shown a fitted sinusoid that peaks at the peak of the interference pattern between $I_{ME}$ and $I_{MV}$.

**Figure 22.** Shows the percentage correlation between the ~ 289SSAN component and a sinusoid peaking at the beat maxima of the interference pattern of $I_{ME}$ and $I_{MV}$. Also shown the variation of SSAN.

**Figure 23.** Compares the averaged SSAN and SSAS spectrum for solar cycles 16-23 (full line) with the spectrum obtained from a similar length series of red noise (broken line).

**Figure 24.** The percentage correlation between the ~88 day period band pass filtered component of red noise and the 88 day period tidal effect due to Mercury.